\documentclass[twocolumn,amssymb, aps, prd, nofootinbib, superscriptaddress, reprint]{revtex4-1}%
\usepackage[utf8]{inputenc}
\usepackage{graphicx}
\usepackage{amssymb}
\usepackage{url,hyperref}
\usepackage{siunitx}
\usepackage{mathtools}
\usepackage{amsmath}
\usepackage{eucal}
\usepackage{xcolor}
\usepackage{lineno}
\usepackage[version=4]{mhchem}
\usepackage{datetime2}
\usepackage{float}

\usepackage[bottom=3.5cm, left=2.cm, right=2 cm]{geometry}

\begin{document}

\title{Quenching factor measurements of neon nuclei in neon gas}
\author{L.~Balogh}
\affiliation{Department of Physics, Engineering Physics \& Astronomy, Queen's University, Kingston, Ontario K7L 3N6, Canada}
\author{C.~Beaufort}
\affiliation{LPSC, Universit\'{e} Grenoble-Alpes, CNRS/IN2P3, Grenoble, France}
\author{A.~Brossard}
\affiliation{Department of Physics, Engineering Physics \& Astronomy, Queen's University, Kingston, Ontario K7L 3N6, Canada}
\author{J.-F.~Caron}
\affiliation{Department of Physics, Engineering Physics \& Astronomy, Queen's University, Kingston, Ontario K7L 3N6, Canada}
\author{M.~Chapellier}
\affiliation{Department of Physics, Engineering Physics \& Astronomy, Queen's University, Kingston, Ontario K7L 3N6, Canada}
\author{J.-M.~Coquillat}
\affiliation{Department of Physics, Engineering Physics \& Astronomy, Queen's University, Kingston, Ontario K7L 3N6, Canada}
\author{E. C.~Corcoran}
\affiliation{Chemistry \& Chemical Engineering Department, Royal Military
College of Canada, Kingston, Ontario K7K 7B4, Canada}
\author{S.~Crawford}
\affiliation{Department of Physics, Engineering Physics \& Astronomy, Queen's University, Kingston, Ontario K7L 3N6, Canada}
\author{A.~Dastgheibi Fard}
\affiliation{LPSC, Universit\'{e} Grenoble-Alpes, CNRS/IN2P3, Grenoble, France}
\author{Y.~Deng}
\affiliation{Department of Physics, University of Alberta, Edmonton, Alberta, T6G 2R3, Canada}
\author{K.~Dering}
\affiliation{Department of Physics, Engineering Physics \& Astronomy, Queen's University, Kingston, Ontario K7L 3N6, Canada}
\author{D.~Durnford}
\affiliation{Department of Physics, University of Alberta, Edmonton, Alberta, T6G 2R3, Canada}
\author{C.~Garrah}
\affiliation{Department of Physics, University of Alberta, Edmonton, Alberta, T6G 2R3, Canada}
\author{G.~Gerbier}
\affiliation{Department of Physics, Engineering Physics \& Astronomy, Queen's University, Kingston, Ontario K7L 3N6, Canada}
\author{I.~Giomataris}
\affiliation{IRFU, CEA, Universit\'{e} Paris-Saclay, F-91191 Gif-sur-Yvette, France}
\author{G.~Giroux}
\affiliation{Department of Physics, Engineering Physics \& Astronomy, Queen's University, Kingston, Ontario K7L 3N6, Canada}
\author{P.~Gorel}
\affiliation{SNOLAB, Lively, Ontario, P3Y 1N2, Canada}
\author{M.~Gros}
\affiliation{IRFU, CEA, Universit\'{e} Paris-Saclay, F-91191 Gif-sur-Yvette, France}
\author{P.~Gros}
\affiliation{Department of Physics, Engineering Physics \& Astronomy, Queen's University, Kingston, Ontario K7L 3N6, Canada}
\author{O.~Guillaudin}
\affiliation{LPSC, Universit\'{e} Grenoble-Alpes, CNRS/IN2P3, Grenoble, France}
\author{E.~W.~Hoppe}
\affiliation{Pacific Northwest National Laboratory, Richland, Washington 99354, USA}
\author{I.~Katsioulas}
\affiliation{School of Physics and Astronomy, University of Birmingham, Birmingham
B15 2TT United Kingdom}
\author{F.~Kelly}
\affiliation{Chemistry \& Chemical Engineering Department, Royal Military
College of Canada, Kingston, Ontario K7K 7B4, Canada}
\author{P.~Knights}
\affiliation{School of Physics and Astronomy, University of Birmingham, Birmingham
B15 2TT United Kingdom}
\author{L.~Kwon}
\affiliation{Chemistry \& Chemical Engineering Department, Royal Military
College of Canada, Kingston, Ontario K7K 7B4, Canada}
\author{S.~Langrock}
\affiliation{SNOLAB, Lively, Ontario, P3Y 1N2, Canada}
\author{P.~Lautridou}
\affiliation{SUBATECH, IMT-AtlantiqueCNRS-IN2P3/Universit\'e de Nantes, Nantes 44307, France}
\author{R. D.~Martin}
\affiliation{Department of Physics, Engineering Physics \& Astronomy, Queen's University, Kingston, Ontario K7L 3N6, Canada}
\author{I.~Manthos}
\affiliation{School of Physics and Astronomy, University of Birmingham, Birmingham
B15 2TT United Kingdom}
\author{J.~Matthews}
\affiliation{School of Physics and Astronomy, University of Birmingham, Birmingham
B15 2TT United Kingdom}
\author{J.-P.~Mols}
\address{IRFU, CEA, Universit\'{e} Paris-Saclay, F-91191 Gif-sur-Yvette, France}
\author{J.-F.~Muraz}
\affiliation{LPSC, Universit\'{e} Grenoble-Alpes, CNRS/IN2P3, Grenoble, France}
\author{T.~Neep}
\affiliation{School of Physics and Astronomy, University of Birmingham, Birmingham
B15 2TT United Kingdom}
\author{K.~Nikolopoulos}
\affiliation{School of Physics and Astronomy, University of Birmingham, Birmingham
B15 2TT United Kingdom}
\author{P.~O'Brien}
\affiliation{Department of Physics, University of Alberta, Edmonton, Alberta, T6G 2R3, Canada}
\author{M.-C.~Piro}
\affiliation{Department of Physics, University of Alberta, Edmonton, Alberta, T6G 2R3, Canada}
\author{P.~Samuleev}
\affiliation{Chemistry \& Chemical Engineering Department, Royal Military
College of Canada, Kingston, Ontario K7K 7B4, Canada}
\author{D.~Santos}
\affiliation{LPSC, Universit\'{e} Grenoble-Alpes, CNRS/IN2P3, Grenoble, France}
\author{G.~Savvidis}
\affiliation{Department of Physics, Engineering Physics \& Astronomy, Queen's University, Kingston, Ontario K7L 3N6, Canada}
\author{I.~Savvidis}
\affiliation{Aristotle University of Thessaloniki, Thessaloniki, Greece}
\author{F.~Vazquez de Sola Fernandez}
\affiliation{SUBATECH, IMT-AtlantiqueCNRS-IN2P3/Universit\'e de Nantes, Nantes 44307, France}
\author{M.~Vidal}
\email{Corresponding author: marie.vidal@queensu.ca}
\affiliation{Department of Physics, Engineering Physics \& Astronomy, Queen's University, Kingston, Ontario K7L 3N6, Canada}
\author{R.~Ward}
\affiliation{School of Physics and Astronomy, University of Birmingham, Birmingham
B15 2TT United Kingdom}
\author{M.~Zampaolo}
\affiliation{LPSC, Universit\'{e} Grenoble-Alpes, CNRS/IN2P3, Grenoble, France}
\collaboration{NEWS-G Collaboration}

\author{P.~An}
\affiliation{Department of Physics, and Triangle Universities Nuclear Laboratory, Duke University, Durham, NC 27708, USA.}
\author{C.~Awe}
\affiliation{Department of Physics, and Triangle Universities Nuclear Laboratory, Duke University, Durham, NC 27708, USA.}
\author{P.~Barbeau}
\affiliation{Department of Physics, and Triangle Universities Nuclear Laboratory, Duke University, Durham, NC 27708, USA.}
\author{S.~Hedges}
\affiliation{Department of Physics, and Triangle Universities Nuclear Laboratory, Duke University, Durham, NC 27708, USA.}
\author{L.~Li}
\affiliation{Department of Physics, and Triangle Universities Nuclear Laboratory, Duke University, Durham, NC 27708, USA.}
\author{J.~Runge}
\affiliation{Department of Physics, and Triangle Universities Nuclear Laboratory, Duke University, Durham, NC 27708, USA.}

\date{\today}

\begin{abstract}

The NEWS-G collaboration uses Spherical Proportional Counters (SPCs) to search for weakly interacting massive particles (WIMPs). In this paper, we report the first measurements of the nuclear quenching factor in neon gas at \SI{2}{bar} using an SPC deployed in a neutron beam at the TUNL facility. The energy-dependence of the nuclear quenching factor is modelled using a simple power law: $\alpha$E$_{nr}^{\beta}$; we determine its parameters by simultaneously fitting the data collected with the detector over a range of energies. We measured the following parameters in Ne:CH$_{4}$ at \SI{2}{bar}: $\alpha$ = 0.2801 $\pm$ 0.0050 (fit) $\pm$ 0.0045 (sys) and $\beta$ = 0.0867 $\pm$ 0.020 (fit) $\pm$ 0.006(sys). Our measurements do not agree with expected values from SRIM or Lindhard theory. We demonstrated the feasibility of performing quenching factor measurements at sub-keV energies in gases using SPCs and a neutron beam.

\end{abstract}
\thispagestyle{empty}
\maketitle

\section{Introduction}
\indent One of the main challenges in particle astrophysics is the search for dark matter. This effort has largely focused on a favored class of low-mass particle candidates called Weakly Interacting Massive Particles (WIMPs) \cite{wimps}. The direct detection of such particles relies on the detection of nuclear recoils of a few keV generated by elastic scatterings on a target nucleus. The detection of Coherent Elastic Neutrino-Nucleus Scattering (CE$\nu$NS) \cite{Freedman_1974} to probe neutrino \cite{drukier}, nuclear \cite{nuclear} and new physics \cite{NSI} also relies on low-energy nuclear recoils.\\
New Experiments with Spheres-Gas (NEWS-G) is an experiment that uses Spherical Proportional Counters (SPCs) to search for WIMPs \cite{newsg} \cite{Giomataris}. The collaboration is also interested in using this technology to detect CE$\nu$NS. In SPCs, the signal of both direct dark matter detection and CE$\nu$NS consists of nuclear recoils from elastic scatters of either dark matter or neutrinos. SPC detectors have appealing features for light dark matter searches and CE$\nu$NS detection, such as sub-keV sensitivity allowed by a high amplification gain and a low intrinsic electronic noise, due to the low capacitance of the sensing electrode. Despite, offering lower interaction cross sections as compared to heavier nucleides, SPCs can be operated with light noble gases (helium, neon) in order to maximize the transfer of momentum from the incoming low mass particle (i.e. $<$ few GeV/c$^2$) to the recoiling nucleus.\\
\indent The energy calibration of detectors is generally achieved using gamma sources inducing electronic recoils in the target material. However, the number of ionized atoms from a nuclear recoil and an electronic recoil of the same energy is different; the nuclear recoil appears ``quenched'' in comparison to the electronic recoil, due to energy dissipation via other channels. We introduce the observed nuclear recoil energy, E$_{ee}$, in electron volts electron equivalent, eV$_{ee}$, as the nuclear recoil energy that is measured by ionization, and the total kinetic nuclear recoil energy, E$_{nr}$, expressed in eV$_{nr}$.The ratio between the number of ionized atoms from a nuclear recoil and an electronic recoil is called the nuclear ionization yield, or nuclear quenching factor (QF). This energy-dependent quantity is essential in understanding the sensitivity of the detector to nuclear recoils. Many quenching factor measurements have been done for semi conductor and scintillator detectors \cite{Mei}. The collaboration for the EDELWEISS dark matter experiment \cite{edelweiss2} has measured the QF in Ge between 20 and \SI{100}{keV} \cite{edelweiss2}, while the CDMS collaboration, also searching for dark matter \cite{CDMS}, measured the QF in Si between 7 and \SI{100}{keV}. The COHERENT collaboration, focusing on studying CE$\nu$NS \cite{coherent}, has measured the QF for CsI[Na] between 5 and \SI{30}{keV}. However, few quenching factor measurements have been done on gas mixtures: $^4$He and isobutane \cite{santos}, \cite{isobutane}. To the best of our knowledge, there are no existing quenching factor measurements in pure neon or a mixture of neon and methane, the latest being the primary gas mixture used by the NEWS-G collaboration. To investigate nuclear recoil energies in the region of interest for a \SI{0.5}{GeV} to few GeV WIMP mass, a suitable neutron energy beam producing sub-keV recoils is necessary.\\
\indent In order to understand the response of SPCs to nuclear recoils in Ne + CH$_4$ (\SI{2}{\%}), used for NEWS-G dark matter searches, we organized a measurement campaign at the Triangle Universities Nuclear Laboratory facility (TUNL) in February 2019. This work presents the first measurement of QF in Ne+CH$_4$ (\SI{3}{\%}) gas for nuclear recoil energies from about 0.34 up to \SI{6.8}{keV_{nr}}, representing the mean of the nuclear recoil distributions. \\
The paper is structured as follows: the experimental set-up is described in Section \ref{section2}. Section \ref{section3} presents the analysis methodology and finally Section \ref{section4} discusses the results.

\section{Experimental Set-up}\label{section2}

\begin{figure}[ht]
\includegraphics[width=.48\textwidth]{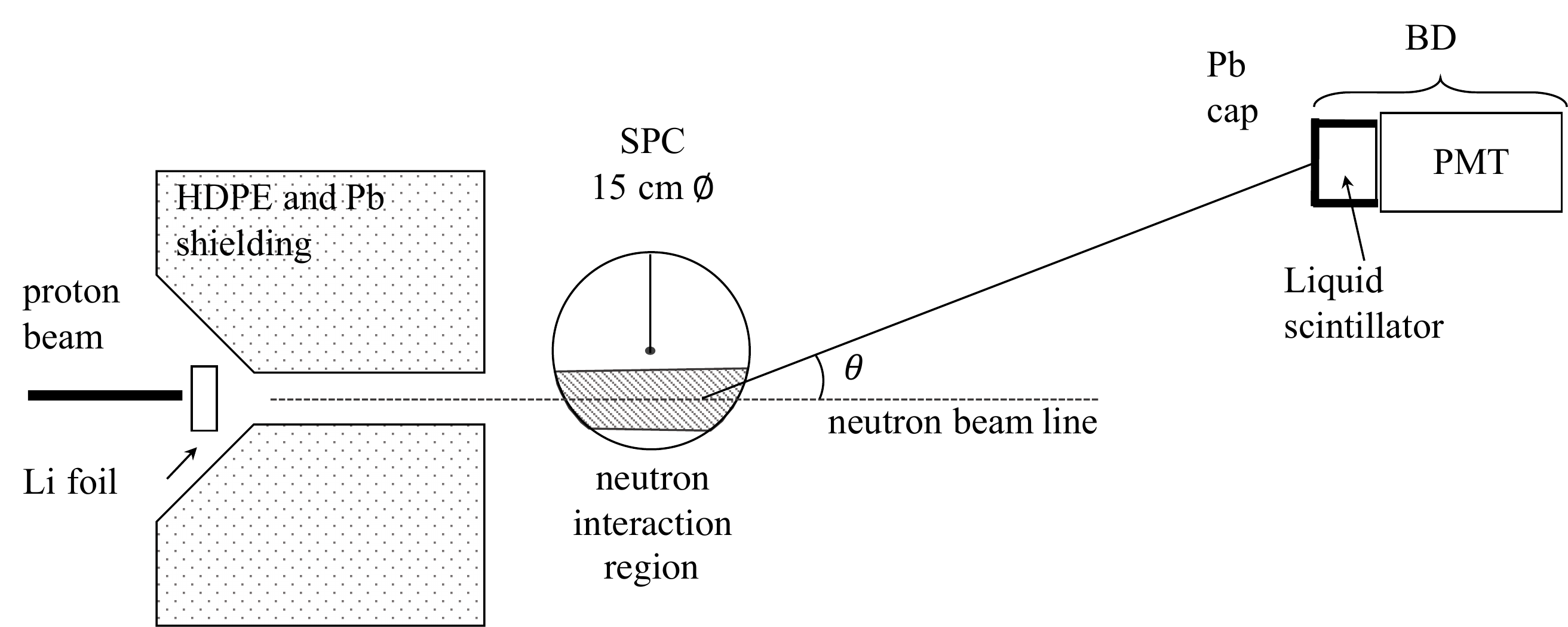}
\caption{\label{tab:scheme_setup}Scheme of the experiment, not at scale. The protons interact with the Li target, generating a neutron beam. A neutron scatters off a nucleus, in the ``south hemisphere" (shaded region), with a scattering angle $\theta$ and is detected by a backing detector (BD).}
\end{figure}

\indent The experimental method was as follows: a target detector is exposed to a neutron beam and the scattered neutrons are recorded by backing detectors (BDs), see Fig. \ref{tab:scheme_setup}. \\
The target detector consists of a \SI{15}{cm} diameter SPC produced by the NEWS-G collaboration. The neutron beam, the backing detectors and the data acquisition system were provided by the TUNL facility. The various elements of the experiment are described in this section.

\begin{figure}[ht]
\includegraphics[width=.34\textwidth]{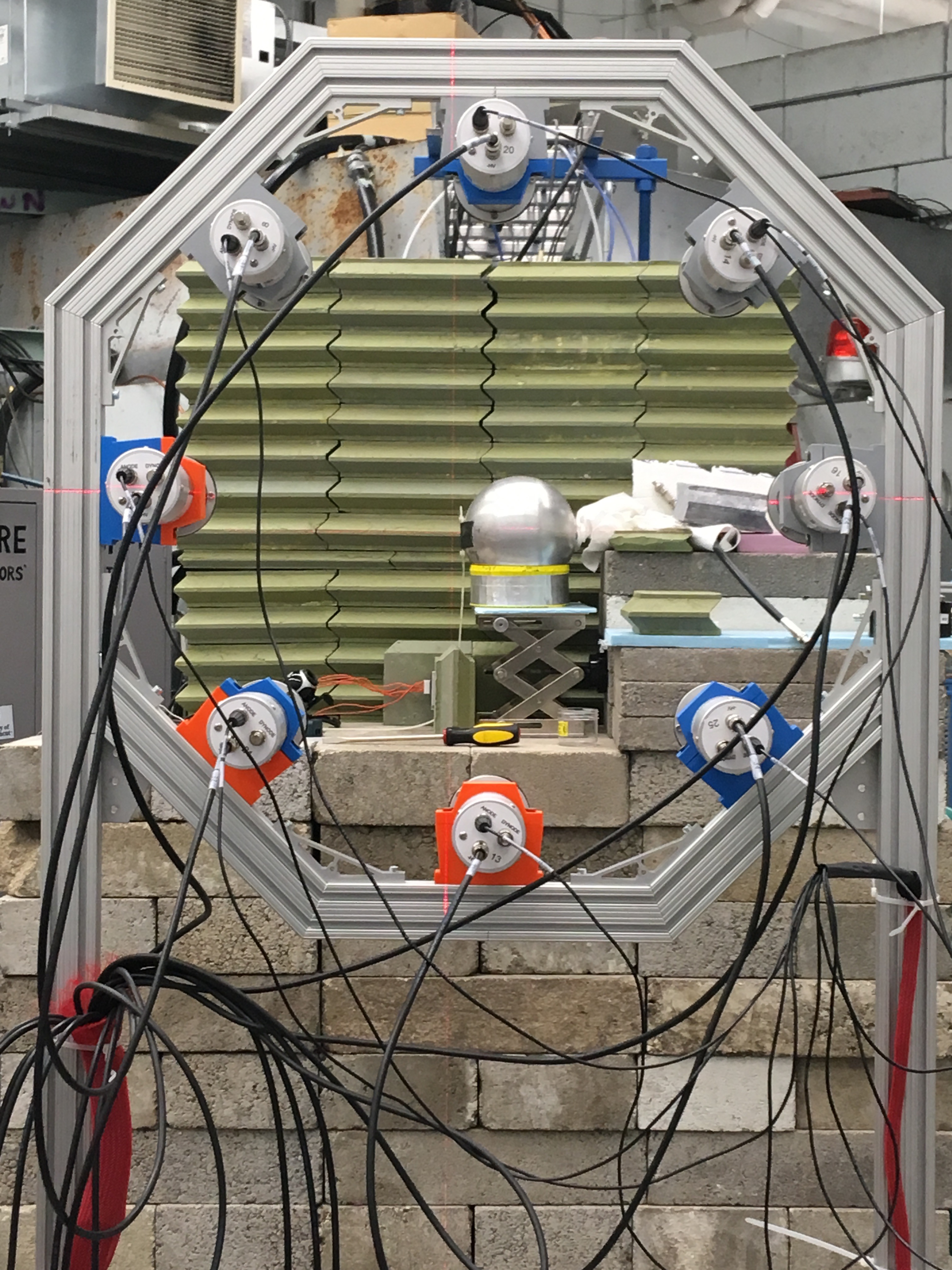}
\caption{\label{tab:photo_setup}Picture of the experimental setup: the SPC is in the center of the photo and the BDs arranged on an annulus structure.}
\end{figure}

\subsection{Spherical Proportional Counter (SPC): the S15 detector}

\indent SPCs were proposed and developed since 2006 at the CEA Saclay (Commissariat \`a l'\'Energie Atomique et aux \'Energies Alternatives) \cite{Giomataris}. This technology consists of a grounded spherical metallic vessel filled with gas (e.g. He, Ne, Ar, CH$_4$). At its center, a small spherical electrode a few mm in size, held by a rod, is set to high voltage (HV) through a wire inside the rod. The electric field generated by the electrode drops off  as $1/r^2$. The volume of the SPC has a large region where the electric field is low (few V/cm), where the e-/ion drift, and a small region where the electric field is high (hundreds of $\mu$m) in the vicinity of the anode and triggers an amplification process via a Townsend avalanche \cite{townsend}.\\
\indent The diffusion of the primary electrons in the drift region has a direct impact on the distribution of their arrival times in the high field region \cite{details_SPC}. Thus, discrimination between surface events associated with high rise times and volume events which have smaller rise times is possible. 

\begin{figure}
\includegraphics[width=.35\textwidth]{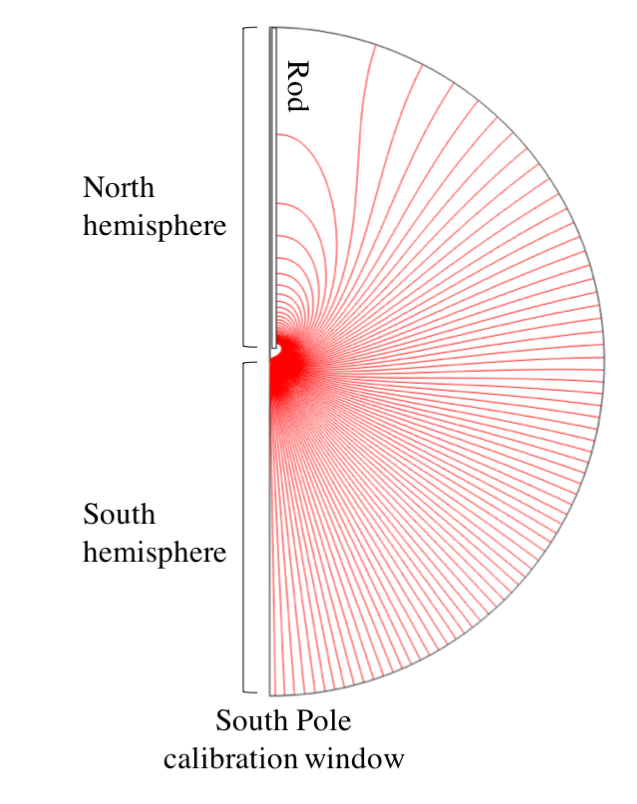}
\caption{\label{Efield}Electric field lines for a spherical proportional counter with a simple sensor design \cite{Paco} \cite{comsol}. The calibration window is placed on the South Pole of the SPC, with a radius of \SI{1}{mm}. Only half of the SPC is shown due to symmetry.}
\end{figure}

A \SI{15}{cm} diameter aluminum SPC with \SI{3}{mm} thick walls was filled with 2 bar of a Ne + CH$_4$ (\SI{3}{\%}) gas mixture. The sensor was a \SI{2}{mm} diameter metallic ball, set to a positive HV of \SI{1700}{V}. The signals were read out using a Canberra 2006 pre-amplifier capacitively coupled to the sensing electrode. With such a simple sensor design (no electric field corrector \cite{details_SPC} \cite{sensor}), the electric field of the ``north hemisphere", containing the rod, is expected to suffer inhomogeneities. Fig. \ref{Efield} shows the electric field lines in our SPC. The neutron beam was aimed at a portion of the volume in the southern hemisphere where the field is homogeneous, see Fig. \ref{tab:scheme_setup}.
To allow for energy calibration with low energy X-rays from a $^{55}$Fe source, the SPC shell was modified to create a thin window, at the south pole location, see Fig. \ref{Efield}.

\subsection{Neutron beam}

\indent Protons from the \SI{20}{MeV} Van De Graaf accelerator at the TUNL facility were used to produce neutrons. Negative hydrogen isotopes are extracted from an ion source to produce a \SI{400}{ns} periodic pulsed proton beam \cite{Howell}. The beam is focused onto a  target made of a \SI{700}{nm} lithium fluoride (LiF) layer on a tantalum foil which produces a monochromatic neutron beam at a given angle through the reaction: 
\begin{equation}
\ce{p + ^7Li -> n + ^7Be + \gamma}
\label{tab:Li}
\end{equation}

A Beam Pickoff Monitor (BPM) was used to identify when the pulsed proton beam interacted in the Li target region, and thus, when neutrons were produced. The neutron beam energy was determined using the difference in time of flight (TOF) between neutrons and gammas, from the lithium target to a liquid-scintillator based backing detector. The neutron energy was determined to be 545 $\pm$ \SI{20}{keV} at a forward angle of zero degrees, from the distribution of the TOF measurements. 

\begin{table*}[ht]
			\begin{center}
				\begin{tabular}{ c c c c c c c }
					\hline
					Run & E$_\text{nr}$ [keV$_\text{nr}$] & $\sigma_{E_{nr}}$ [keV$_\text{nr}$] &  $\theta$ [$^o$] & $\sigma_{\theta}$ [$^o$] & Distance [cm] & Exposure\\ [0.5ex]
					\hline
					8 & 6.80 & 1.15 & 29.02 $\pm$ 0.4 & 2.45 & 44.6 $\pm$ 0.4 & 4h\\
					7 & 2.93 & 0.46 & 18.84 $\pm$ 0.1 & 1.47 & 77.9 $\pm$ 0.2  & 7h14\\
					14 & 2.02 & 0.29 & 15.63 $\pm$ 0.3 & 1.12 & 103.4 $\pm$ 1.6 & 36h21\\
					9 & 1.70 & 0.26 & 14.33 $\pm$ 0.06 & 1.1 & 106.8 $\pm$ 0.1 & 16h\\
					10 & 1.30 & 0.2 & 12.48 $\pm$ 0.05 & 0.94 & 124.7 $\pm$ 0.1 & 23h\\
					14 & 1.03 &  0.2 & 11.13 $\pm$ 0.3 & 1.1 & 103.7 $\pm$ 1.5 & 36h21\\
					11 & 0.74 & 0.11 & 9.4 $\pm$ 0.03 & 0.69 & 169.3 $\pm$ 0.08 & 33h22\\
					14 & 0.34 & 0.11 & 6.33 $\pm$ 0.26 & 1.1 & 104.4 $\pm$ 0.5 & 36h21\\[1ex]
					\hline
				\end{tabular}
				\caption{\label{tab:table_winter}Table with the chosen nuclear recoil energies, their corresponding scattering angles (from measurements taken on site), the distance from the SPC to the BDs and time exposures. Run 7, 8, 9, 10 and 11 were recorded with the annulus structure, while run 14 recorded three nuclear recoil energies simultaneously.}
			\end{center}
		\end{table*}

\subsection {Scattered neutron detectors}

\indent Scattered neutrons were recorded by ``backing" detectors (BDs) consisting of a liquid scintillator EJ-309 cell, from Eljen \cite{eljen}, coupled to a Hamamatsu R7724 photomultiplier tubes \cite{pmt}. The properties of this liquid scintillator allowed neutrons and gammas to be differentiated by pulse shape discrimination (PSD). The PSD method relies on different ionization density, which produces different scintillation signals characteristics. \\
\indent Knowing the incident neutron energy and the scattering angle, the nuclear recoil energy deposited in the SPC can be determined through kinematics \cite{Enr}:
\begin{equation}
\begin{split}
E_{nr}&(\theta_s, E_n) = 2E_n \dfrac{M_n^2}{(M_n + M_T)^2} \times \\
&\Bigg( \dfrac{M_T}{M_n} + \sin^2 \theta_s - \cos \theta_s \sqrt{\bigg(\dfrac{M_T}{M_n}\bigg)^2-\sin^2\theta_s}  \Bigg),
\end{split}
\end{equation}
where $\theta_s$ is the scattering angle of the neutron with respect to its initial trajectory, $E_n$ is the incident neutron energy, $M_n$ is the neutron mass and $M_{T}$ is the target mass of the nucleus.\\

Eight BDs were arranged on an annulus structure with a radius of \SI{29.4}{cm} allowing multiple backing detectors to record neutrons for a given scattering angle, see Figure \ref{tab:photo_setup}. By changing the distance between the SPC and annulus structure, this configuration allowed us to vary the scattering angle and record nuclear recoil energies ranging from \SI{6.8}{keV_{nr}} down to \SI{0.74}{keV_{nr}}, see Table \ref{tab:table_winter}.

A second configuration was adopted to reach the lowest energy recoils in order to increase statistics. We used three pairs of BDs placed at about \SI{1}{m} from the SPC and set at three different scattering angles. The smaller scattering angle, of \SI{6.3}{^o}, allowed us to reach a desired \SI{0.34}{keV_{nr}} nuclear recoil energy. The two additional energies overlapped energies taken with the annulus structure during this campaign, see Table \ref{tab:table_winter}. To reduce the number of  gamma rays  and undesired scattered neutrons interacting in the SPC, the lithium target and beam line (collimator) were shielded with lead and high-density polyethylene and the backing detectors were shielded from gammas with \SI{2}{mm} thick lead caps.\\

\begin{figure*}
\centering
\begin{minipage}[b]{.45\textwidth}
\includegraphics[width=1.1\textwidth]{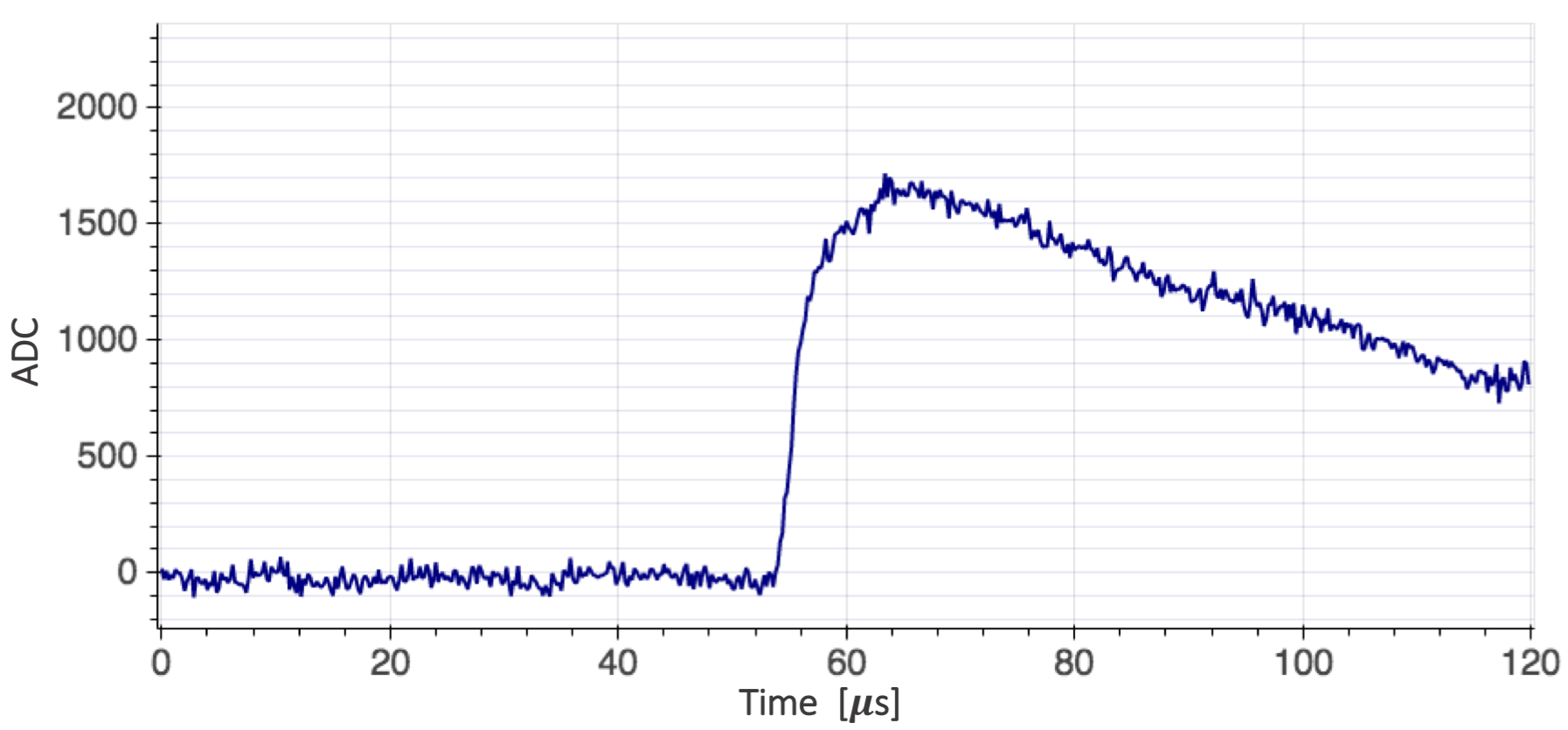}
\caption{Example of SPC raw pulse.}\label{raw}
\end{minipage}\qquad
\begin{minipage}[b]{.45\textwidth}
\includegraphics[width=1.05\textwidth]{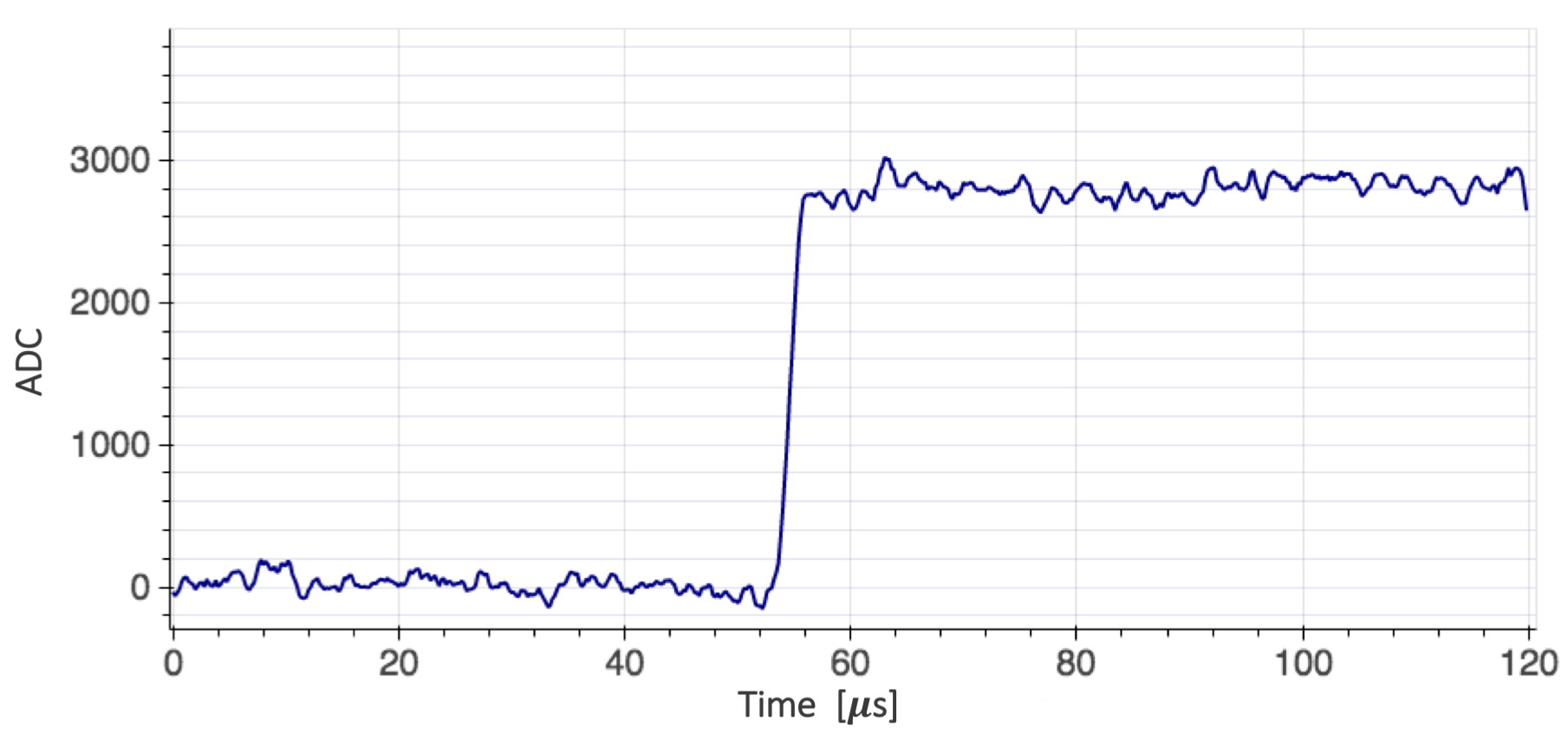}
\caption{Example of SPC treated pulse.}\label{treated}
\end{minipage}
\end{figure*}

\subsection {Data acquisition and processing}	

\indent The Data Acquisition System (DAQ) was set up to trigger on a single BD, which would trigger a SIS3316 digitizer to record traces from all of the BDs, the SPC and the BPM, sampled at 250 MHz. Because the DAQ triggers on the BDs and not on the SPC, we recorded SPC signals without an energy threshold.\\
\indent The SPC pulses were treated by de-convolving the electronic response of the pre-amplifier and the ion drift from the recorded signal \cite{Paco}. Fig. \ref{raw} and \ref{treated} show an example of raw pulse and the resulting treated pulse respectively. This method corrects for the ballistic deficit effect on the amplitude and the rise time of the pulses. The rise time gives an estimation of the diffusion of the primary electrons along their drift toward the anode and thus of the radial distance of the event. The rise time is calculated between 10 and \SI{90}{\%} of the pulse amplitude. 
The recorded quantities determined from the recorded signals and used in the analysis are the following :
\begin{itemize}
    \item[--] Amplitude of the pulse, from the deconvolved digitized SPC pulse: corresponds to the energy estimator for the event.
    \item[--] Rise time of the pulse, from the deconvolved digitized SPC pulse,
    \item[--] ``Onset time": the time between the BD pulse (trigger of the DAQ) and the SPC pulse. It is defined as the time between the interaction and the start of the pulse, i.e. the drift time of the electrons to the anode minus the TOF between the interaction time  and the BD. Since the TOF of the scattered neutrons from the SPC to the BDs is of the order of a few hundred nanoseconds, while the drift time of the primary electrons in the SPC is of the order of tens of $\mu s$, the onset time is essentially equal to the drift time of electrons in the SPC.
    \item[--] BPM time: the time between the BD pulse and the BPM, measures the total TOF and allows for events outside the beam pulse window to be rejected,
    \item[--] The neutron/gamma pulse shape discrimination parameter, PSD, calculated  from digitized BD pulse. It is the ratio of the total integrated pulse to the integrated charge in the tail of the pulse.
\end{itemize}

\begin{figure*}
\centering
\begin{minipage}[b]{.39\textwidth}
\includegraphics[width=1.1\textwidth]{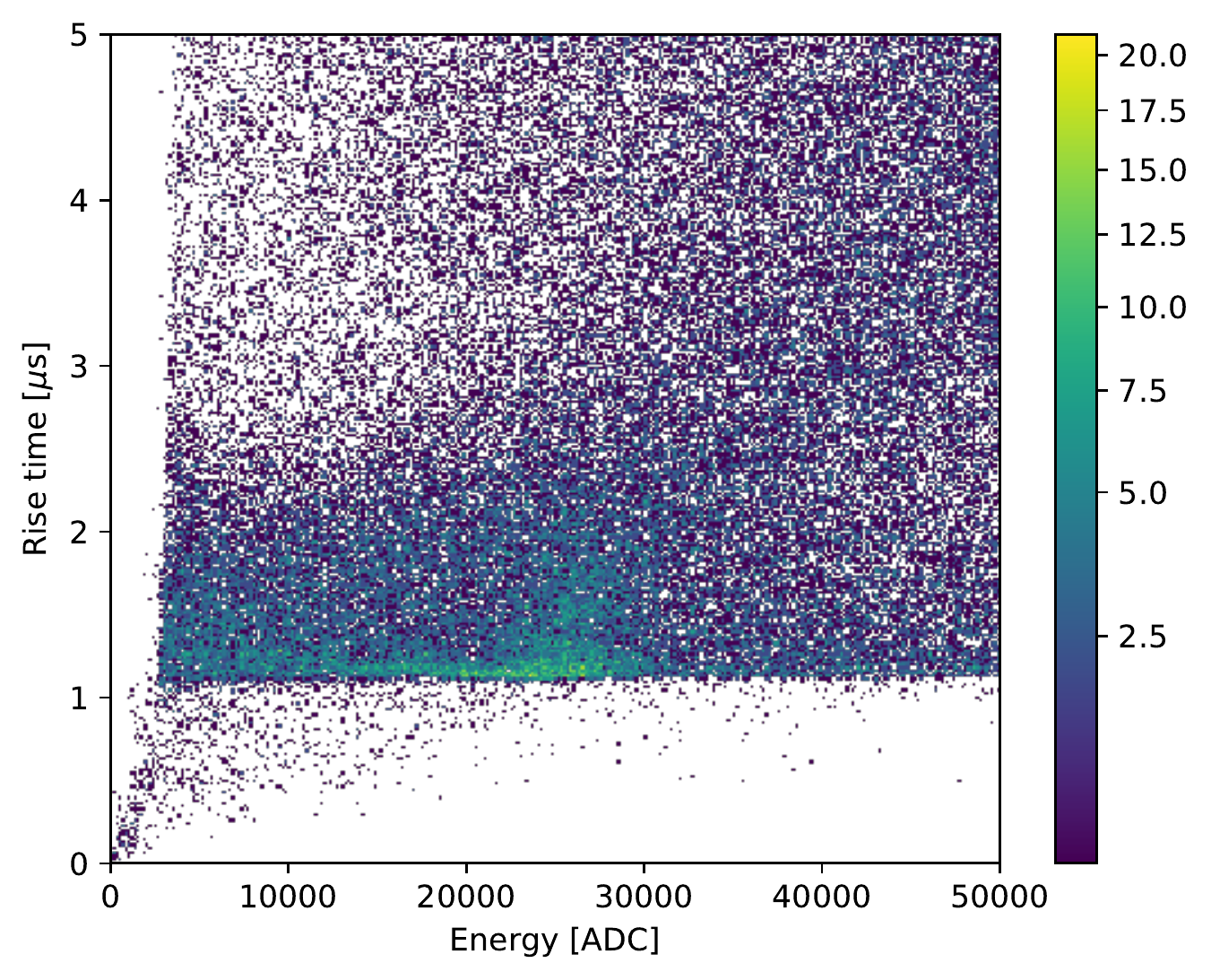}
\caption{$^{55}$Fe data with beam ON. Rise time as a function of energy of the events. The population of events at \SI{25000}{ADC} and between 1 and \SI{1.51}{\mu s} corresponds to $^{55}$Fe events, see text for more details.}\label{Fe55rt}
\end{minipage}\qquad
\begin{minipage}[b]{.42\textwidth}
\includegraphics[width=1.1\textwidth]{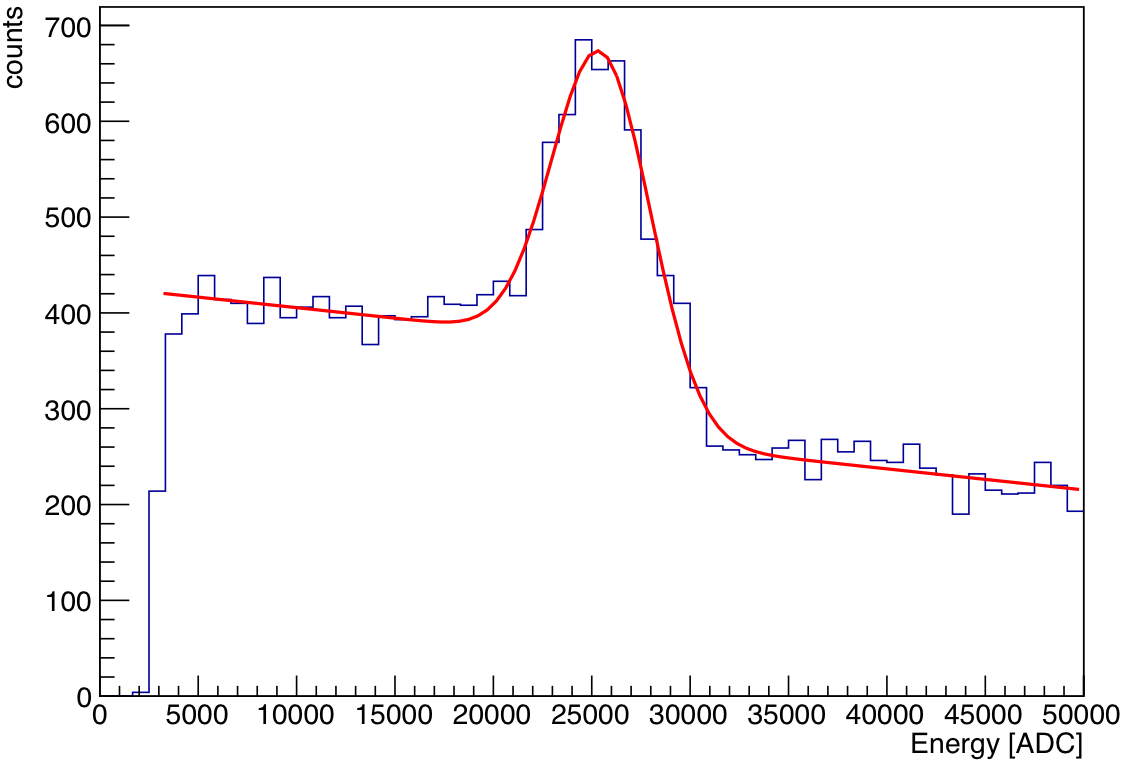}
\caption{$^{55}$Fe energy spectrum extracted from rise time cuts: between 1 and \SI{1.51}{\mu s}. The red curve is the fit to the data using the model described in the text. The fit allowed to extract the energy scale of the experiment.}\label{Fe55spectrum}
\end{minipage}
\end{figure*}

The DAQ configuration was set up with a trigger delay of \SI{40}{\mu s}. It was applied to the SPC traces in order to center the traces within the \SI{120}{\mu s} pulse recording window used for the SPC data. 

Details of the runs are given in Table \ref{tab:table_winter}.  

\subsection{Energy calibration and gain monitoring}

\begin{figure}
\includegraphics[width=.48\textwidth]{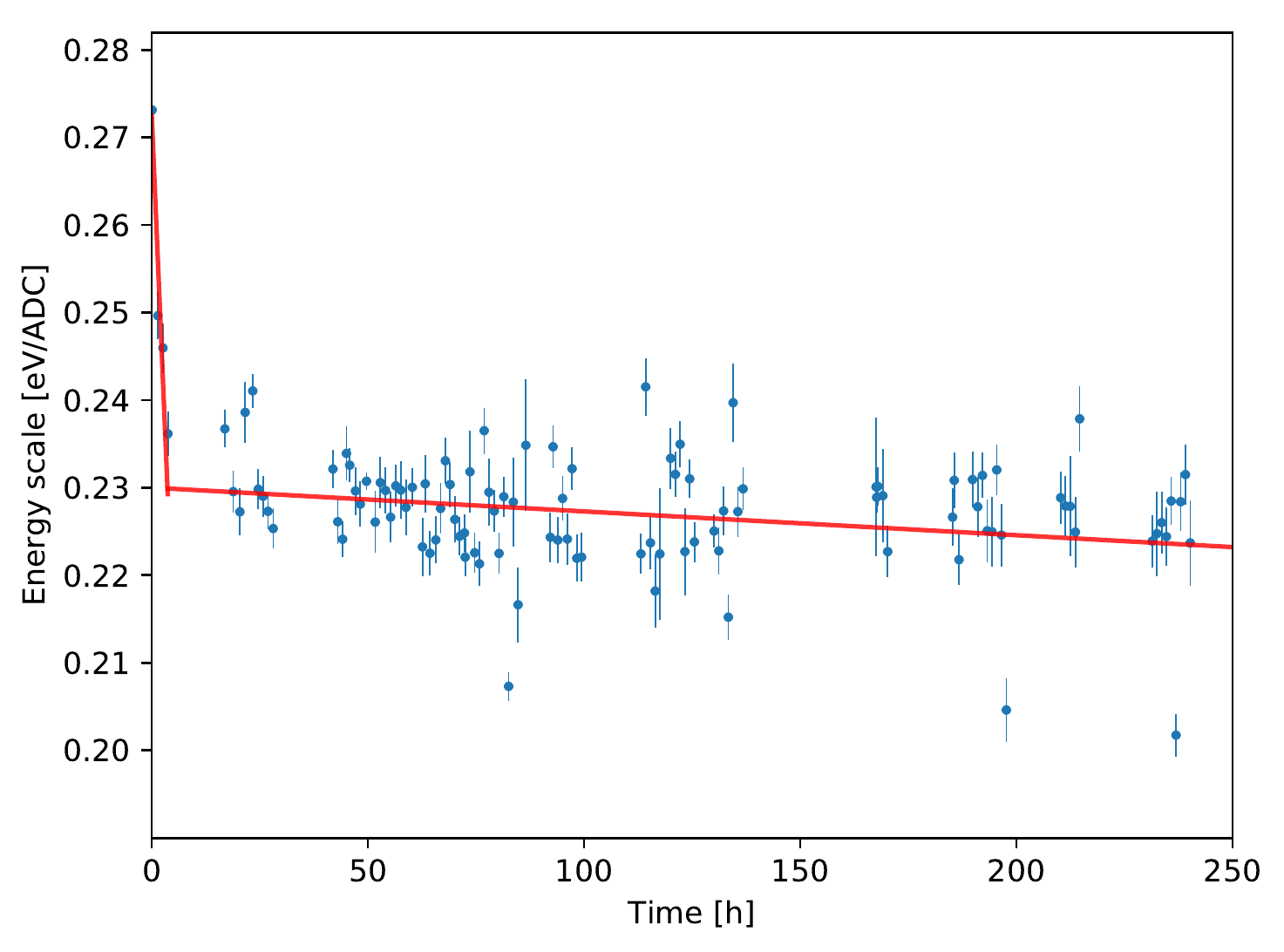}
\caption{\label{tab:energy_scale}Energy scale determined by the $^{55}$Fe calibration source as a function of time: blue data points. We observe that the detector reached stability after 4 hours of data taking. The stability of the detector was further studied by looking at background events monitoring the evolution of the gain for the entirety of the experiment. This study confirmed the trend hereby shown by the $^{55}$Fe energy scale and that the number of outliers represent \SI{0.83}{\%} of all the data. These data were then fitted with a piecewise linear function, shown in red. }
\end{figure}

\indent The energy calibration of the SPC was carried out in-situ  using an $^{55}$Fe source placed next to the calibration window at the south pole of the SPC. 

The calibration data monitored the gain of the SPC every hour with the beam on. To select $^{55}$Fe events, we applied a strong rise time cut from 1 to \SI{1.51}{\mu s} to reject background events, see Fig. \ref{Fe55rt}. This allowed the \SI{5.9}{keV} peak \cite{Fe55} to be extracted, see Fig. \ref{Fe55spectrum}. We modelled the $^{55}$Fe peak with a Gaussian and the background with a sum of a complementary error function with a linear function for the background, see Fig. \ref{Fe55spectrum}. The mean value of each peak returned by the fit of each calibration data set was extracted and used to monitor and correct the energy scale as a function of time, see Fig. \ref{tab:energy_scale}. The gain changed by 5 percent during the \SI{250}{hours} of data taking.
The energy of the SPC events were then corrected using the fit results of the energy scale as a function of time. The averaged energy conversion factor is \SI{0.23}{eV/ADC}.

The linearity of the energy response of the detector was inferred based on existing calibration data with other SPC detectors. In particular, the energy response in neon gas was confirmed using $^{37}$Ar whose decay produces two monoenergetic lines at \SI{2.82}{keV} and at \SI{270}{eV} in \cite{laser}. A study of a possible non-linearity of the detector response was performed to investigate any impact on our results, it will be detailed in Section \ref{analysis}.

\begin{figure*}
\centering
\begin{minipage}[b]{.45\textwidth}
\includegraphics[width=1.05\textwidth]{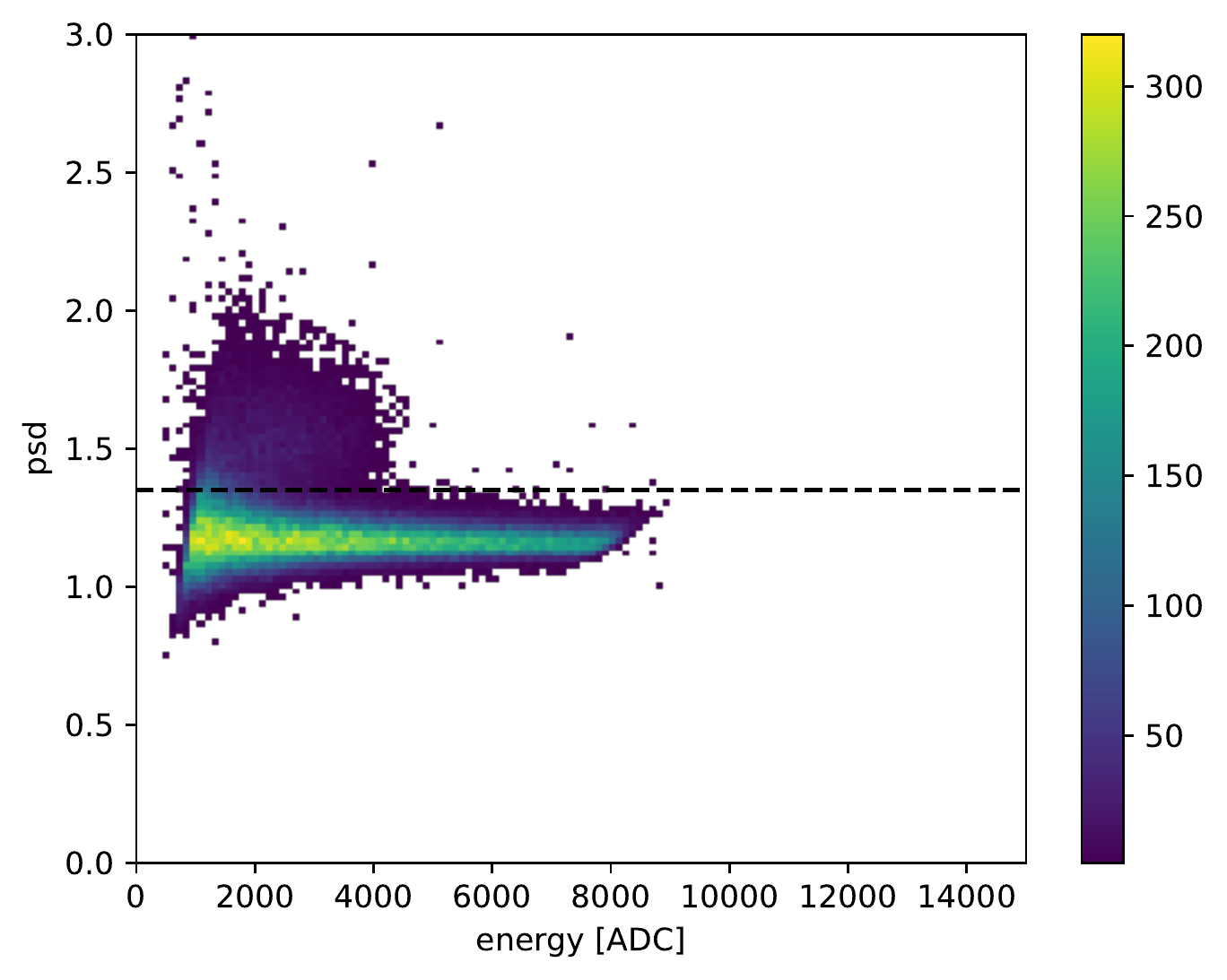}
\caption{PSD as a function of the energy of the events in a backing detector for run8's configuration. We set the PSD cut such that most gamma events would be rejected and neutron events selected.}\label{psd}
\end{minipage}\qquad
\begin{minipage}[b]{.45\textwidth}
\includegraphics[width=1.05\textwidth]{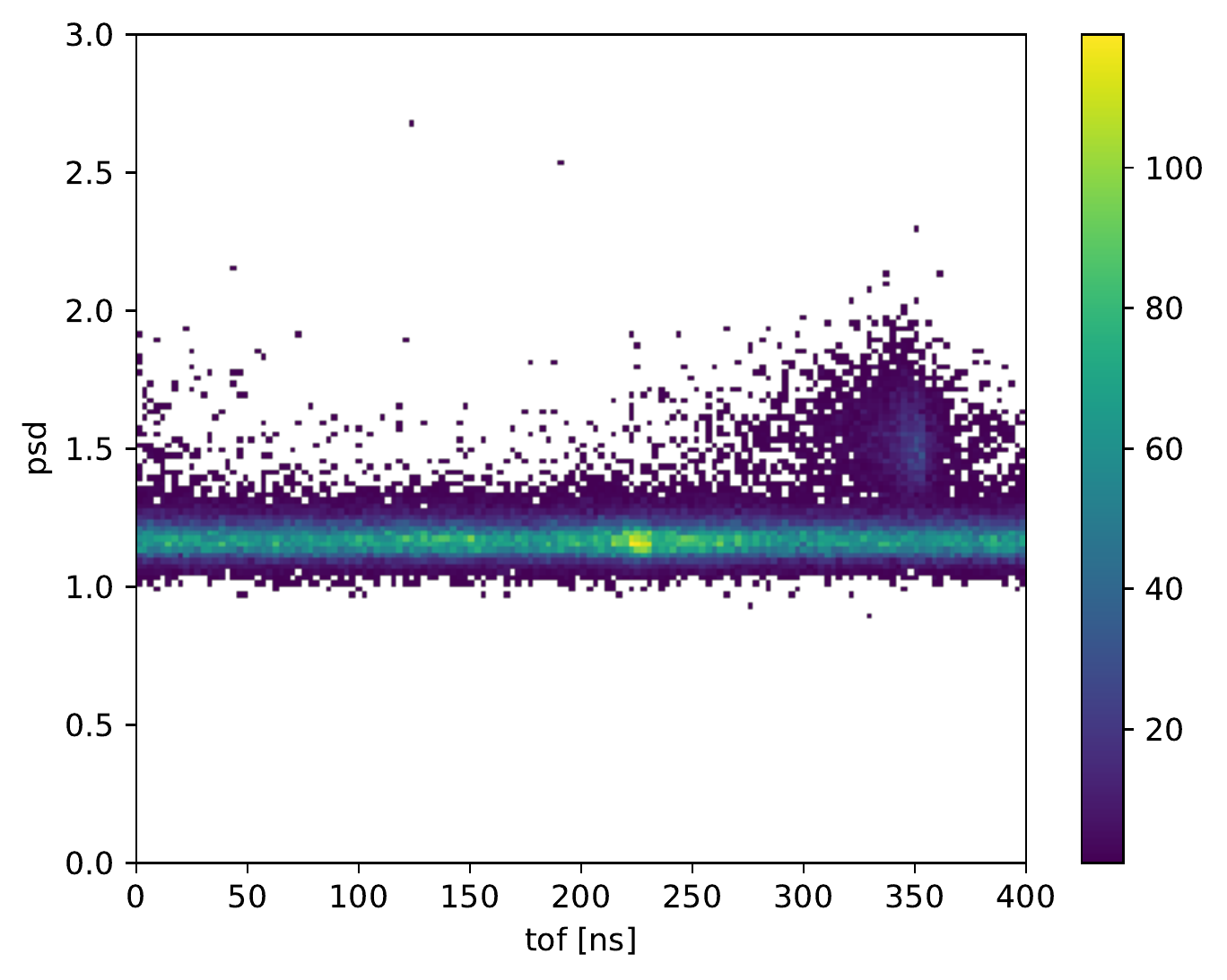}
\caption{PSD as a function of TOF corresponding to run8's configuration. The TOF for gamma events associated with the beam is \SI{225}{ns} and TOF for neutrons is centred at \SI{350}{ns}.} \label{psdvstof}
\end{minipage}
\end{figure*}

\section{Analysis}\label{section3}

\subsection{Neutron run: event selection}
We used several selection cuts to discriminate nuclear recoil events from background events. 
A first set of cuts to select events consistent with neutron interactions in the BDs was performed. Fig. \ref{psd} shows two populations: neutrons centred at a PSD of 1.7 and gammas centred at 1.2. The events with a PSD larger than 1.35 were selected. Second, we kept the events with a TOF consistent with neutrons (target-BD). Fig. \ref{psdvstof} shows the gamma and neutron events associated with the beam as well as gamma events from ambient radiation. The TOF cut varies depending on the annulus configuration, Table \ref{tab:tof} lists the different values. \\

\begin{figure*}
\centering
\begin{minipage}[b]{.3\textwidth}
\includegraphics[width=1.1\textwidth]{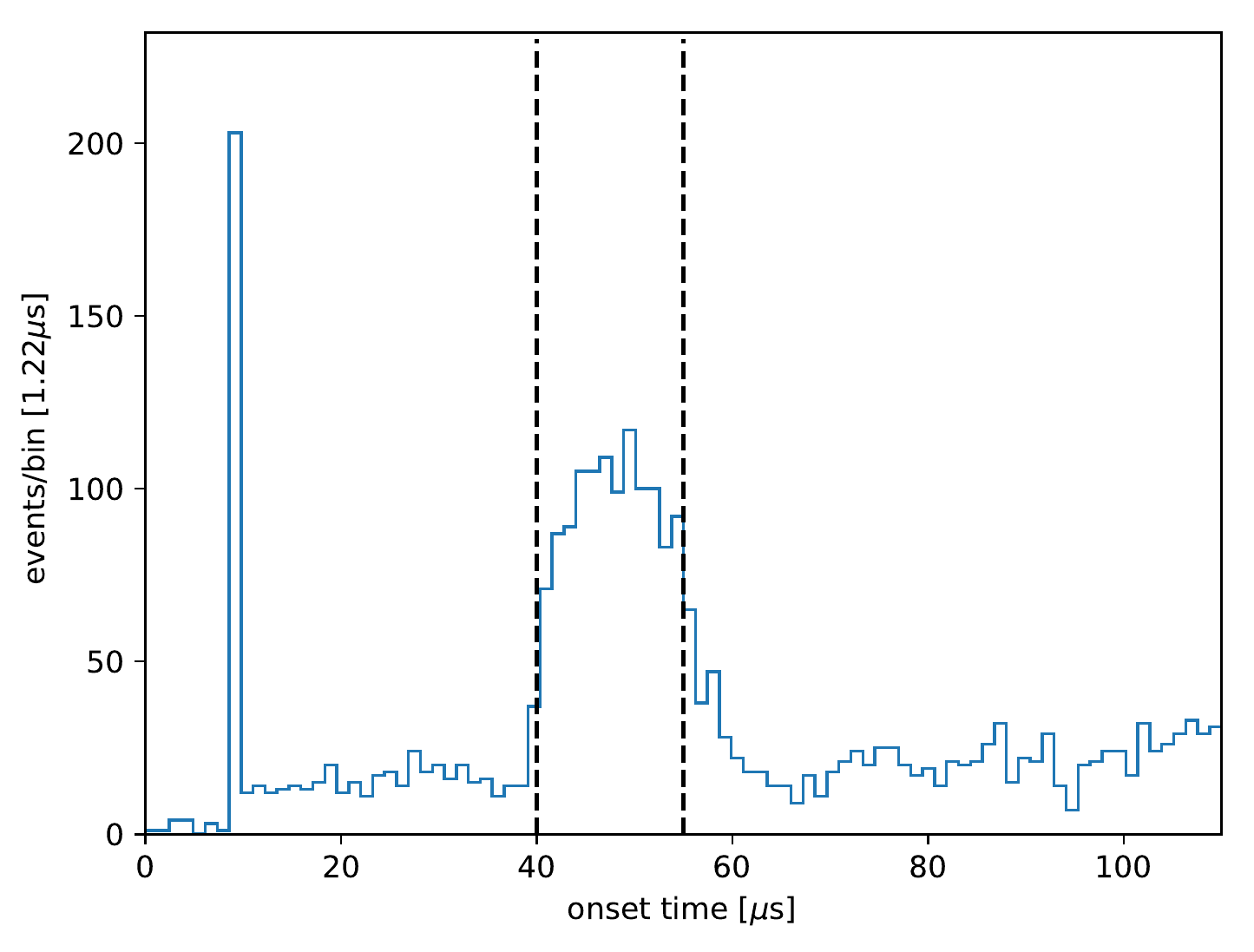}
\caption{Onset time spectrum for run8 after PSD and TOF cuts. Observation of an excess of events between 40 and \SI{55}{\mu s} representing the coincident events between the SPC and the BDs.}\label{onset}
\end{minipage}\qquad
\begin{minipage}[b]{.3\textwidth}
\includegraphics[width=1.1\textwidth]{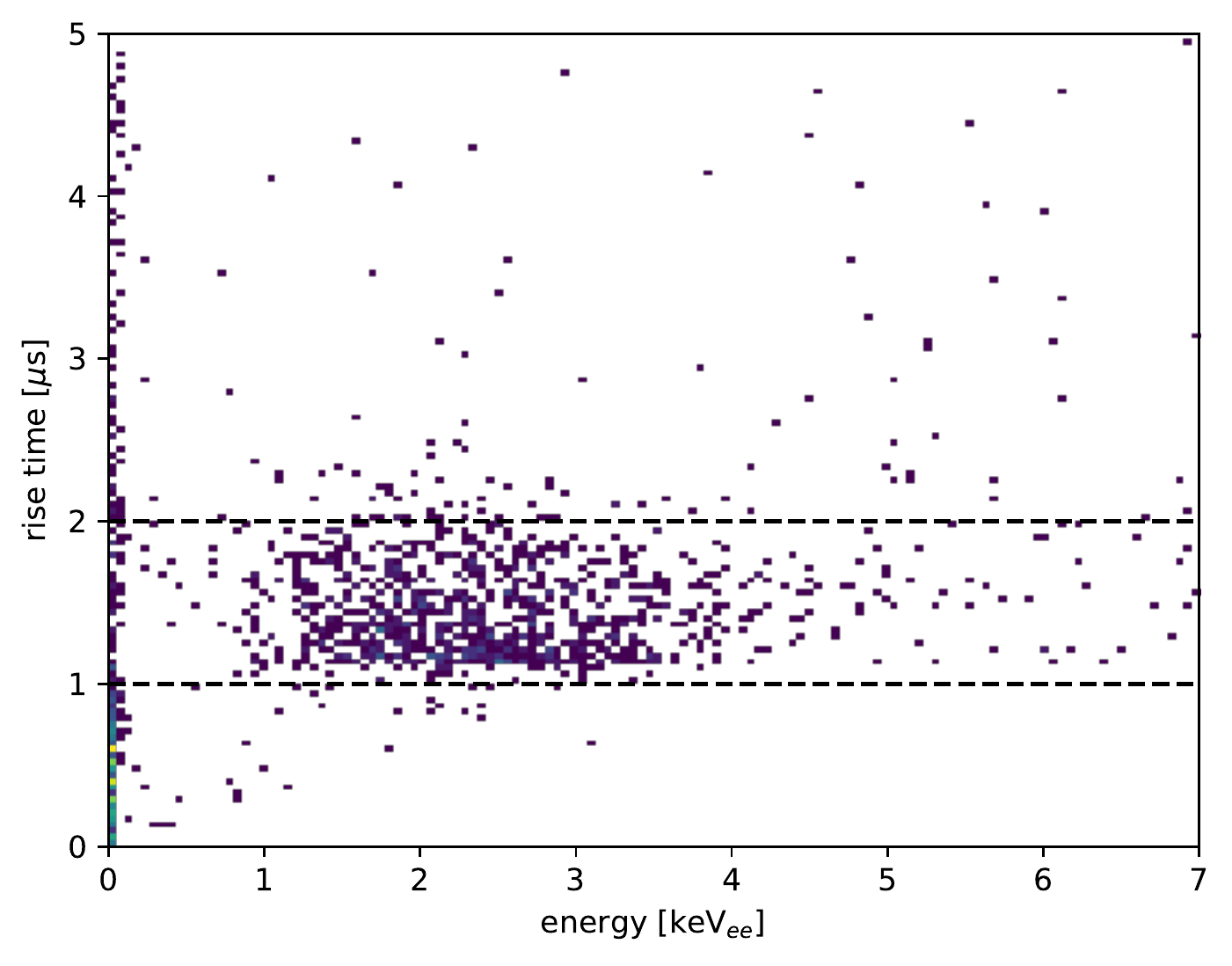}
\caption{Rise time as a function of energy for run8 after PSD, TOF and onset time cuts. The concentration of events represents the neon nuclear recoils. Events between 1 and \SI{2}{\mu s} are selected to build the energy spectra.}\label{rt_energy}
\end{minipage}\qquad
\begin{minipage}[b]{.3\textwidth}
\includegraphics[width=1.1\textwidth]{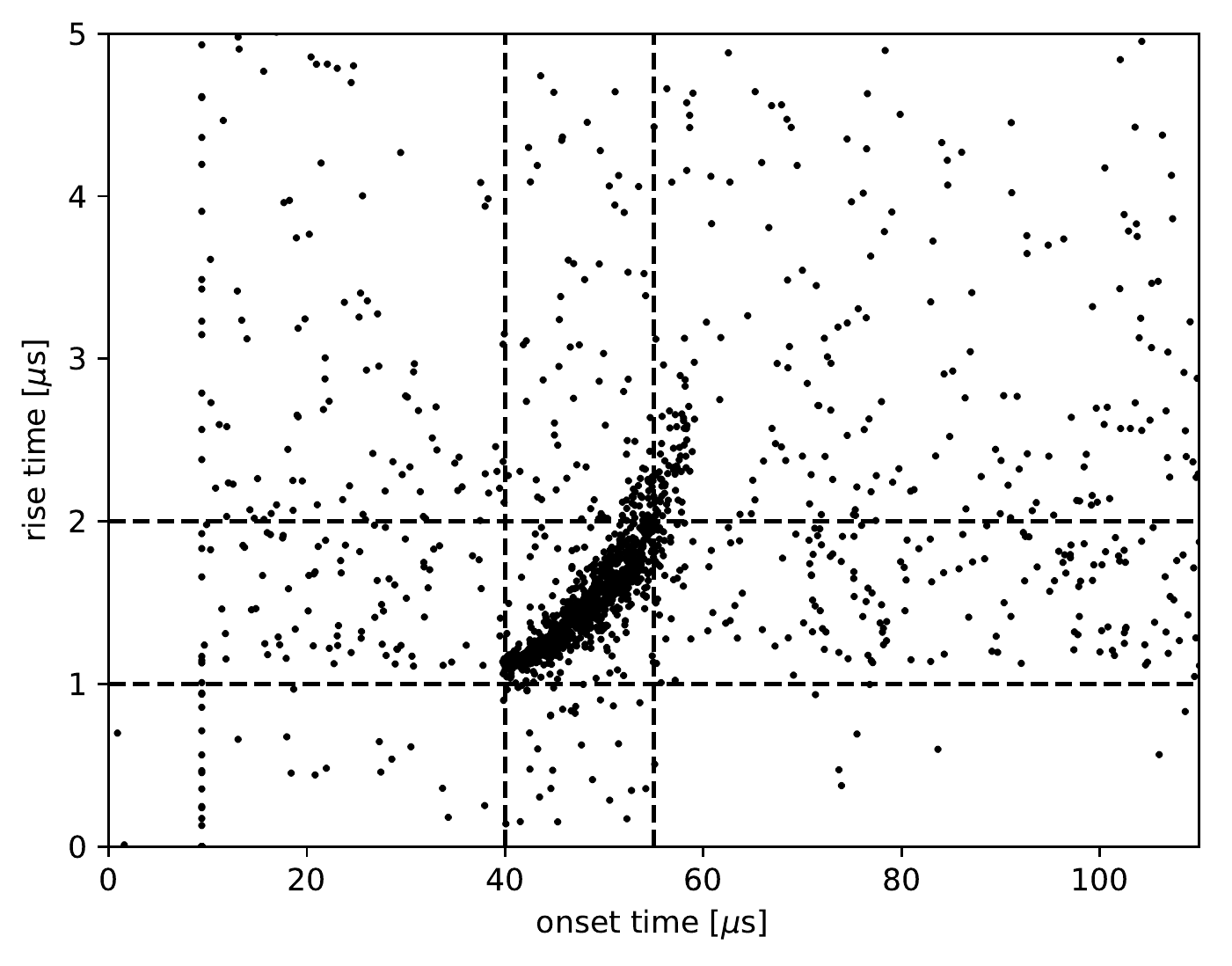}
\caption{Rise time as a function of the onset time (drift time of the events) for run8 after PSD and TOF cuts. The population with correlated rise time and onset time correspond to volume events in the SPC.}\label{rt_onset}
\end{minipage}
\end{figure*}

A second set of cuts are performed on SPC related variables. 
We used the onset time variable to select events consistent with neutrons scattering off of nuclei in the SPC. As described above, the onset time distribution is expected to show an excess of events starting at \SI{40}{\mu s} due to the DAQ configuration. 
Fig. \ref{onset} shows, indeed, an excess of events between 40 and \SI{55}{\mu s} corresponding to nuclear recoils induced by neutrons. The events with an onset time between 40 and \SI{55}{\mu s} were selected.
Fig. \ref{rt_energy} shows the rise time as a function of energy (run8) after the onset time cut. The population of events between 1 and \SI{4}{keV} is consistent with recoil events. Finally, only the interactions in the volume were kept,  with rise times between 1 and \SI{2}{\mu s}. This cut is used to reject track events such as muons (high rise time) and a significant part of the noise (low rise time). 
Fig. \ref{rt_onset} shows the correlation between the rise time of SPC events and their onset time, which is the drift time of the events. This observed feature is characteristic of interactions taking place in the volume of the detector. The cuts in rise time and onset time were determined from the scatter plots of the different runs.  \\
The resulting energy spectra of all runs, after applying all cuts above, are shown in Fig. \ref{tab:results_fit} in Section \ref{section4}.

\begin{table}
\centering
	\begin{tabular}{ c c }
			\hline
			Run number & TOF cut [ns]\\[0.5ex]
			\hline
			run 8  & 95 - 140 \\
			run 7 & 135 - 185 \\
			run 9 & 140 - 205 \\
			run 10 & 150 - 215\\
			run 11 & 205 - 265 \\
			run 14 & 145 - 205 \\[1ex]
			\hline
		\end{tabular}
		\caption{Table summarizing the TOF cut for the different runs. The TOF varies depending on the energy run due to the BDs distance relative to the SPC.\label{tab:tof}}
\end{table}

\subsection{Modelling of the recoil peak}

\indent The rather large spread in nuclear recoil energies, the non Gaussian shape, and the energy dependence of the quenching factor precludes the more traditional method consisting in estimating the nuclear response by computing the ratio of the energy spectrum mean with the nuclear recoil energy obtained from kinematics. \\
\indent The shape of the recoil distribution depends on several factors, including the geometry of the SPC and BDs, the energy of the neutrons, the quenching factor, and other effects. In this Section, we derive the model used to describe the expected recoil distribution. \\

\indent The width of the recoil peak is strongly affected by the scattering angle distribution, which is determined by the geometry of the experiment. The beam had a cross section of $5\times$\SI{5}{cm^2} at \SI{2.54}{cm} from the exit of the collimator and $6\times$\SI{6}{cm^2} at \SI{27.94}{cm}. The cross section of the beam is assumed to be square due to the collimator hole having a square shape. The neutron beam cross section was scanned and found to be approximately uniform, thus it was modelled as such. The scan of the beam was performed in two locations along the beam line and looked for differences in the neutron population along the orthogonal plane to the beam line.

\begin{figure}
\includegraphics[width=.48\textwidth]{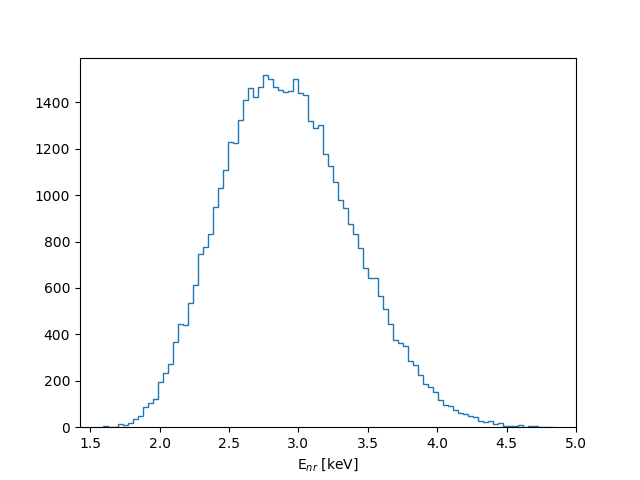}
\caption{\label{tab:Enr} Simulated nuclear recoil energy spectrum for a distance between the annulus structure and the surface of the SPC of \SI{77.9}{cm}, which corresponds to run7. It takes into account the neutron beam energy distribution and the scattering angle distribution (geometry of the experiment).}
\end{figure}

\indent A Monte Carlo simulation recreating the geometry of the experiment allowed us to model the distribution of the scattering angle for each energy run, by simulating the neutron interactions in the SPC and in the BDs.  Fig. \ref{tab:Enr} shows the simulated nuclear recoil energy spectrum for the run with a mean energy of \SI{2.93}{keV} (run7); the shape of the peak is due to the spread in energy of the neutron beam as well as the distribution in scattering angles resulting from the size of the SPC, BDs and neutron beam. \\

\indent Lindhard theory is often used to model energy losses from recoiling nuclei \cite{Lindhard}, measurements in germanium and silicon are consistent with the theory \cite{cogent} \cite{damic} \cite{supercdms}. The model consists of 11 parameters and is energy dependent.\\
We implemented a QF that varies with the energy as each energy run samples a range of energies where the QF varies. We opted for a simple 2 parameters parametrization, used by the EDELWEISS collaboration \cite{edelweiss}, which matched well their data. This simple parametrization can model the quenching factor calculated by SRIM (Stopping and Range of Ions in Matter) \cite{srim} and the quenching factor from the Lindhard theory. The quenching factor is parametrized as follow:
\begin{equation}\label{par}
QF(E_{nr}) = \alpha E_{nr}^{\beta}.
\end{equation}
This parametrization covers a wide range of shapes depending on the values of $\alpha$ and $\beta$. \\

\indent In order to predict the expected nuclear recoil distribution, we need to model the response of our detector.
\indent The primary ionization electrons created from the recoiling nucleus drift towards the detector anode, resulting in an avalanche of secondary ionization electrons, greatly amplifying the signal \cite{details_SPC}. The primary ionization process is modelled by assuming that the number of primary ionized atoms follows a Poisson distribution, whose mean depends on the scattering angle, the neutron energy and QF. This choice is motivated by the rather broad recoil energy distribution, which include greater dispersion than anticipated for primary ionization in neon with an expected Fano factor of approximately 0.2 \cite{laser} \cite{fano}. The number of secondary electron/ion pairs is modelled using a Polya distribution  \cite{micromegas2000}, \cite{micromegas2009}, \cite{gem2017}, \cite{gem2007}:\\

\begin{equation}
\begin{split}
P_{\text{Polya}}(S) = &\dfrac{1}{\langle G \rangle} \dfrac{(1+\theta_p)^{1+\theta_p}}{\Gamma(1+\theta_p)} \Bigg( \dfrac{S}{\langle G \rangle}\Bigg)^{\theta_p} \\
&\times \exp\Bigg(-(1+\theta_p) \dfrac{S}{\langle G \rangle}\Bigg),
\end{split}
\end{equation}\\
where S is the number of secondary electrons for a single primary electron entering the avalanche region, $\langle G \rangle$ is the mean gain and $\theta_p$ governs the shape of the distribution. Each electron's avalanche is considered independent from the others, thus the probability of creating S secondary electrons given N primary electrons reaching the avalanche region is given by the N$^\text{th}$ convolution of the Polya distribution:
\begin{equation}
\begin{split}
P_{\text{Polya} }^{N^{th}}(S \mid N )  &= \dfrac{1}{\langle G \rangle} \Bigg ( \dfrac{(1+\theta_p)^{(1+\theta_p)}}{\Gamma(1+\theta_p)}\Bigg)^{N}  \Bigg( \dfrac{S}{\langle G \rangle}\Bigg)^{N(1+\theta_p)-1} \\
&\times \exp\Bigg(-(1+\theta_p) \Big( \dfrac{S}{\langle G \rangle}\Big)\Bigg)\\
&\times \prod^{N-1}_{j=1}B(j+j\theta_p, 1+\theta_p)
\end{split}
\end{equation}
where $B(x, y)$ is the Beta function.\\

\indent There are residual fluctuations of the gain throughout the volume of the detector due to the anisotropy of the electric field lines. The volume sampled by the $^{55}$Fe calibration with \SI{5.9}{keV} X-rays is smaller than the volume sampled by the nuclear recoils. A Geant4 simulation \cite{Geant4} showed that about \SI{70}{\%} of $^{55}$Fe events are located in the south hemisphere. Therefore, we include a term that allows the energy scale to fluctuate following a normal distribution, with mean 1, normalized to the ADC/eV conversion factor measured above, and standard deviation $\sigma_a$.\\

\indent The two most abundant isotopes of neon were included in the model: $^{20}$Ne and $^{22}$Ne with \SI{90.48}{\%} and \SI{9.25}{\%} abundance. The interaction rates of neutrons with each isotope was calculated using Geant4, $^{20}$Ne representing \SI{92.4}{\%} of the interactions and $^{22}$Ne \SI{7.6}{\%}. It was assumed that the two isotopes have the same quenching factor and W-value (mean ionization energy necessary to create an electron ion pair). The other isotopes present in the gas mixture were carbon and hydrogen. Using the same Geant4 simulation, the interaction rates on carbon and hydrogen are \SI{1.7}{\%} and \SI{11.9}{\%} respectively. The interactions on these isotopes were not included in the model; the contribution from carbon recoils are negligible in comparison to the neon recoils and the proton recoils take place outside of the energy range covered by the joint fit.\\
Finally, a study of the reconstruction efficiency and rise time cuts was performed and modelled by an error function:
\begin{equation}
\varepsilon (j_{pe}) = a_e \times \mathrm{erf}\Bigg(\dfrac{j_{pe}- b_e}{c_e}\Bigg),
\label{eq:eff}
\end{equation}
where $a_e$, $b_e$ and $c_e$ are coefficients of the model function and $j_{pe}$ the number of primary electrons.\\ From fitting Equation \ref{eq:eff} to the  simulated data to measure the energy reconstruction efficiency, we obtained $a_e$ = 0.874, $b_e$ = -0.12 and $c_e$ = 8.01.\\

The probability density function of getting the energy $x_i$ from a recoil event, is given by:
\begin{equation}
\begin{split}
&P_s(x_i) = \sum^{N_{pe}}_{j_{pe}=1} \Big(\varepsilon(j_{pe}) \int_a P_{\text{Polya}}^{N^{th}}(x_i \mid a, j_{pe}, I)   \\
& \times P_a(a \mid \mu_a, \sigma_a) \int_{\theta_s} \int_{E_n} P_{\text{Poisson}}(j_{pe} \mid \mu_{j_{pe}}(\theta_s, E_n, \alpha, \beta, I)) \\
&\times P_{\theta_s}(\theta_s \mid \mu_{\theta_s}, \sigma_{\theta_s}) P_{E_n}(E_n \mid \mu_{E_n}, \sigma_{E_n})\Big)
\end{split}
\label{eq:ps}
\end{equation}
where $I$ denotes fixed parameters of the model that are not specified: the mean gain: $\langle G \rangle$=1000, the W-value and $\theta_p $: \SI{27.6}{eV} and 0.12 respectively \cite{laser}. $P_a$, $P_{\theta}$ and $P_{E_n}$ are the distributions of the energy scale, the scattering angle and the neutron energy respectively. They are modelled as normal distributions with means: $\mu_a$, $\mu_{\theta_s}$ and $\mu_{E_n}$ and standard deviations: $\sigma_a$, $\sigma_{\theta_s}$ and $\sigma_{E_n}$. $a$ is the energy scale, $E_n$ is the neutron energy, $\theta_s$ is the scattering angle, $N_{pe}$ is the maximum number of primary electrons, $\alpha$ and $\beta$ the parameters of the quenching factor function. The recoil energy spectra from the data can then be fitted using the above Equation \ref{eq:ps}, in order to determine $\alpha$ and $\beta$, and thus the energy dependence of the quenching factor.

\subsection{Analysis approach}\label{analysis}

\begin{figure}
\includegraphics[scale=0.55]{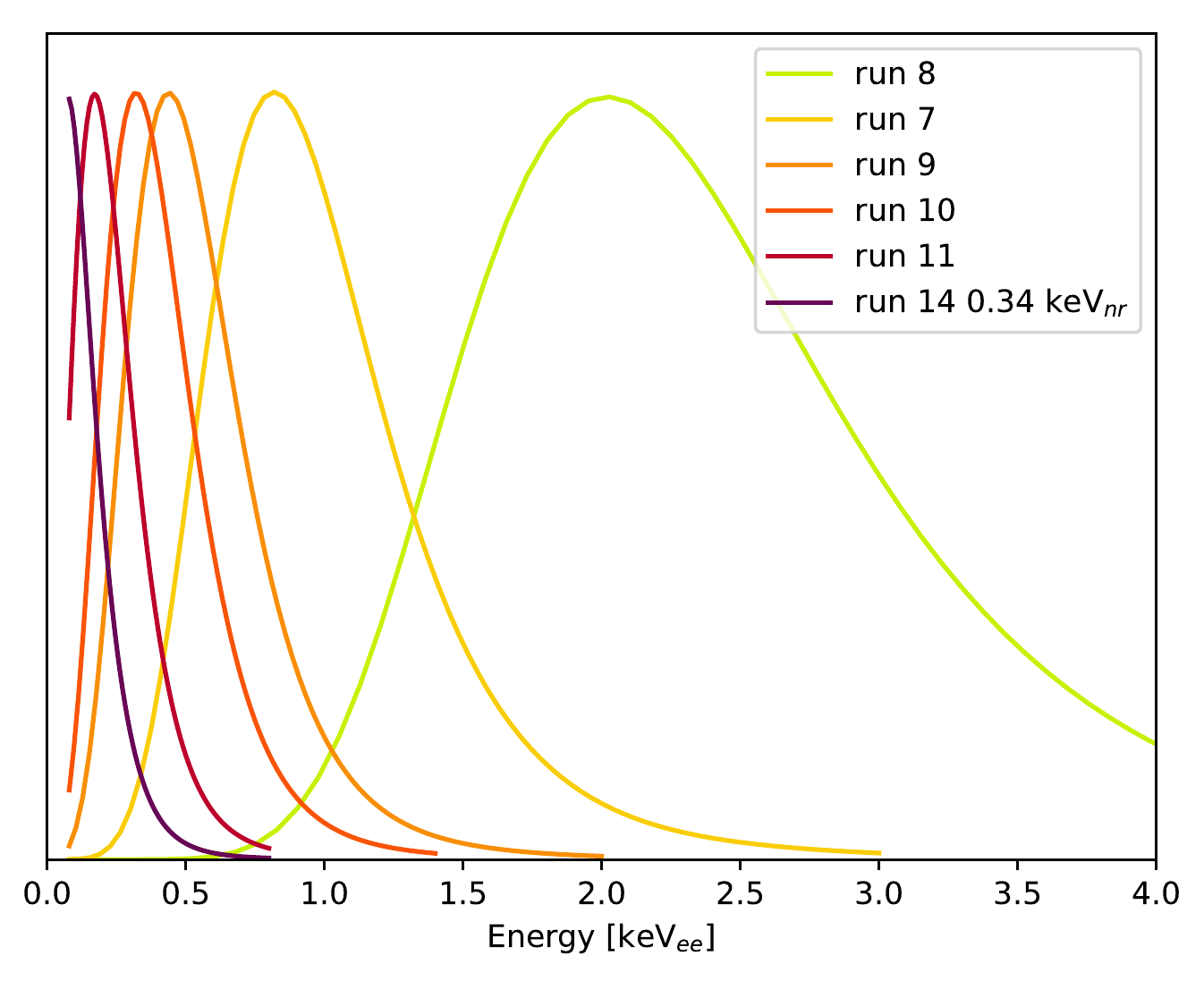}
\caption{\label{tab:overlap_Eee} Energy spectra extracted from the results of the fit for 6 energy runs: 8, 7, 9, 10, 11 and 14 \SI{0.34}{keV_{nr}}, that shows the overlap between the different runs.}
\end{figure}

Because of the overlap in energy from one energy run to another, see Fig. \ref{tab:overlap_Eee}, a joint (or simultaneous) unbinned fit over the data sets from the different run configurations of the model was performed. Hence, the energy ranges common to different energy runs provide an increased statistical accuracy on the quenching factor. The analysis threshold was set to \SI{100}{eV} to avoid the noise events at very low energies. The likelihood function for the number of events in the data as a function of energy was used:

\begin{equation}
-\ln(L)= -\sum^N_i\ln\Big( P_s(x_i) f_s+ P_{BG}(x_i) f_{BG}\Big),
\end{equation}
where $P_s$ is the probability density of getting the energy $x_i$ from a signal event, given in Equation \ref{eq:ps} and $f_s$ is the fraction of expected signal events. $P_{BG}$ is the probability density for the background events and is assumed to be uniform in energy based on the background energy spectra built with the onset time cuts out of the signal window. The events selected are environmental radiation, which gives us a good estimate of our background in the signal window. $f_{BG}$ is the fraction of expected background events. \\
The final function to minimize includes Gaussian priors on the scattering angle means, coming from measurements provided in Table \ref{tab:table_winter}, in order to include systematic uncertainties from our measurements. The posterior for 1 energy run is therefore:
\begin{equation}
\begin{split}
&- \ln(P) = -\sum^N_i  \Bigg[\ln\Bigg(f_s\Big(0.924P_s^{^{20}Ne}(x_i)\\
&+0.076P_s^{^{22}Ne}(x_i) \Big)+ f_{BG}P_{BG}(x_i) \Bigg)- \ln(p_{\theta_s}(\theta_i))\Bigg],
\end{split}
\end{equation}
where $p_{\theta_s}$ is a Gaussian prior on the angle.\\
Finally, the sum of the 8 negative log-posteriors (8 runs) is minimized:
\begin{equation}
\begin{split}
\ln \Big(P(I, f_s, \alpha, &\beta, \theta_s, \sigma_a | x_i)\Big) = \\
&\sum^{runs}_j \ln \Big(P_j(I, f_s, \alpha, \beta, \theta_s, \sigma_a | x_i) \Big)
\end{split}
\end{equation}

The fraction of expected signal events for each run, the parameters of the quenching factor, $\alpha$  and $\beta$, the scattering angle mean for each run, $\theta_s$, and the standard deviation of the energy scale distribution, $\sigma_a$, are free parameters of the fit. The quenching factor parameters and the standard deviation of the energy scale are common to all energy runs, thus constrained by all the data sets. Overall, the fit has 19 free parameters ($f_s$ and $\theta_s$ for each energy run, $\alpha$, $\beta$ and $\sigma_a$) and 6 fixed parameters ($E_n, \sigma_{E_n}, \sigma_{\theta_s}, \theta_p,$ the W-value and the mean gain). $f_s$ and $\sigma_a$ were bounded by limits between 0 and 1.\\
The fit was performed using iminuit \cite{iminuit}, which is a python package based on the Minuit minimization library \cite{minuit}. We fit over different energy ranges depending on the data set, see Table \ref{tab:table_chi}.\\

In addition to fitting all of the energy runs simultaneously, we fit each run independently for comparison. Only the standard deviation of the energy scale distribution (fluctuation of the gain) was fixed because it is a common parameter of the energy runs. We were thus able to extract independent quenching factors for each run and compare them with the joint fit results. This will be discussed in the next section.\\

Table \ref{tab:unc_summary} summarizes the different sources of uncertainties studied. As mentioned before, the systematic uncertainties on the mean scattering angle is accounted for directly in the fit by floating the mean scattering angle. The uncertainties from the neutron energy and the baseline noise fluctuation have a negligible impact on the quenching factor parameters and the quenching factor. The impact from a non-linear response of the detector ($\pm$\SI{0.7}{\%} \cite{laser}) and the efficiency curve were studied and have a small impact on the quenching factor: up to \SI{0.7}{\%} and \SI{0.8}{\%} respectively. The systematic uncertainties from these two sources will be reported decoupled from the other errors in Section \ref{section4}.\\
For the interested reader, an in depth explanation of the analysis and experimental setup is provided in \cite{vidal}.

 \begin{figure*}
\centering
\includegraphics[width=.3\textwidth]{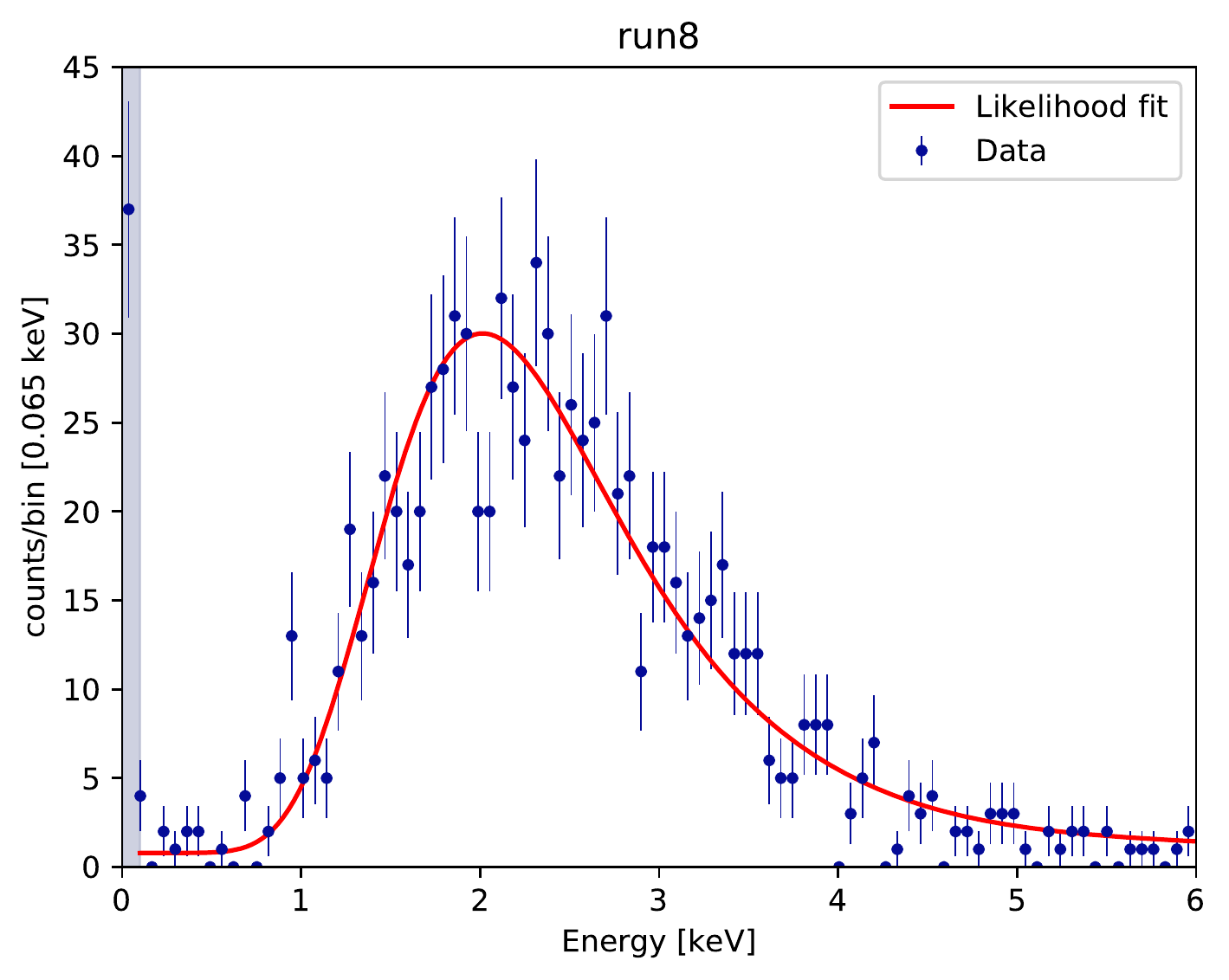}\quad
\includegraphics[width=.3\textwidth]{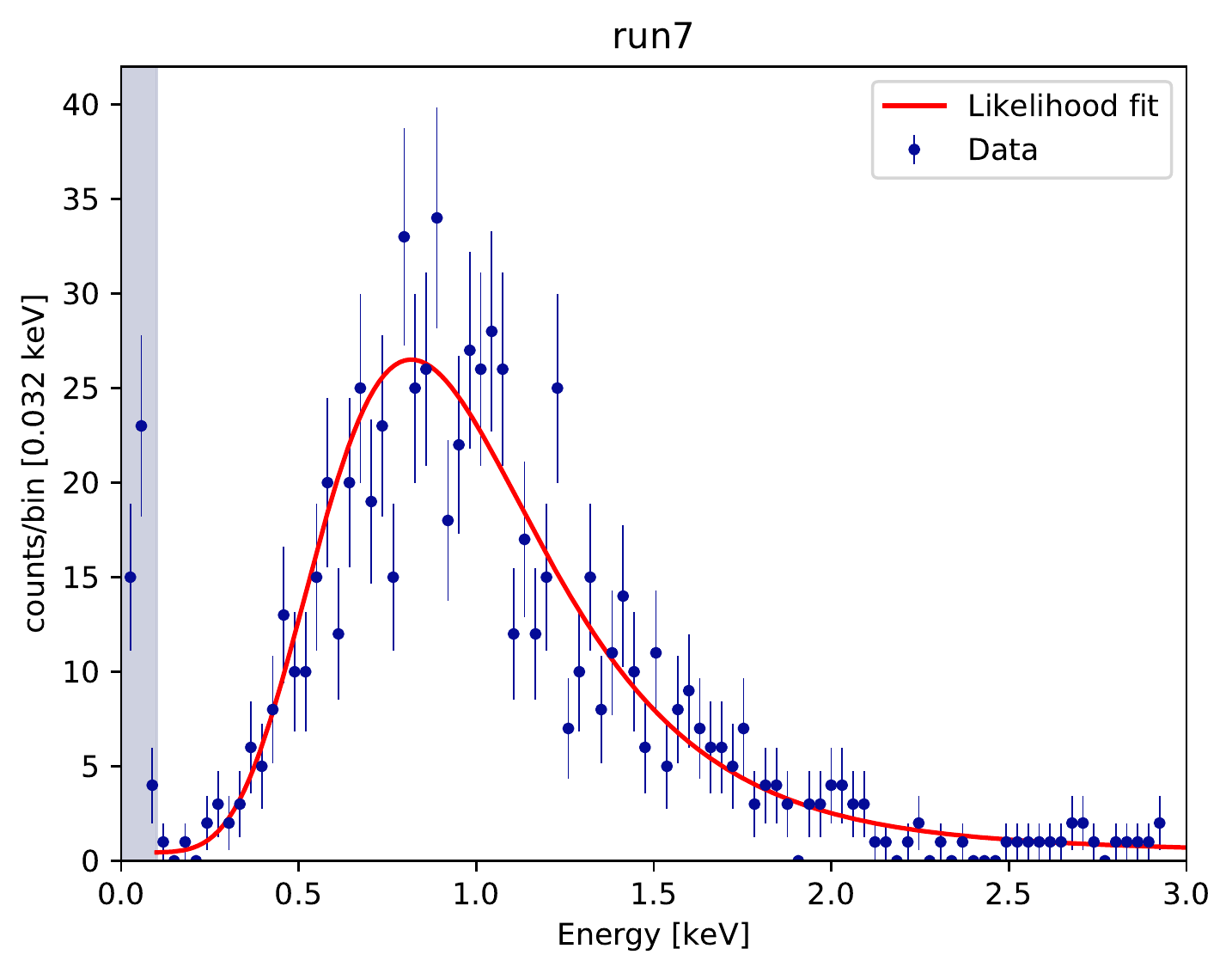}\quad
\includegraphics[width=.3\textwidth]{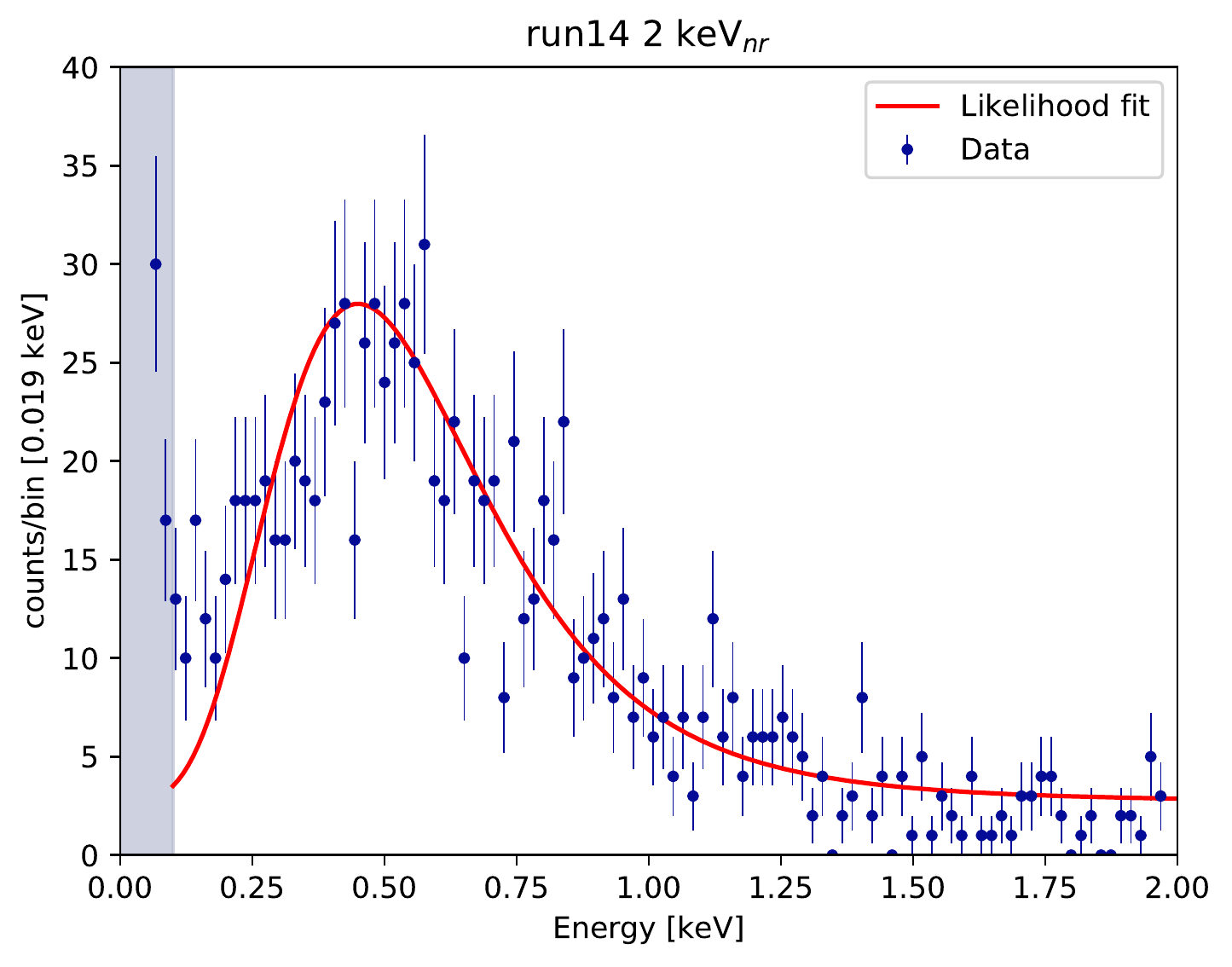}
\medskip
\includegraphics[width=.3\textwidth]{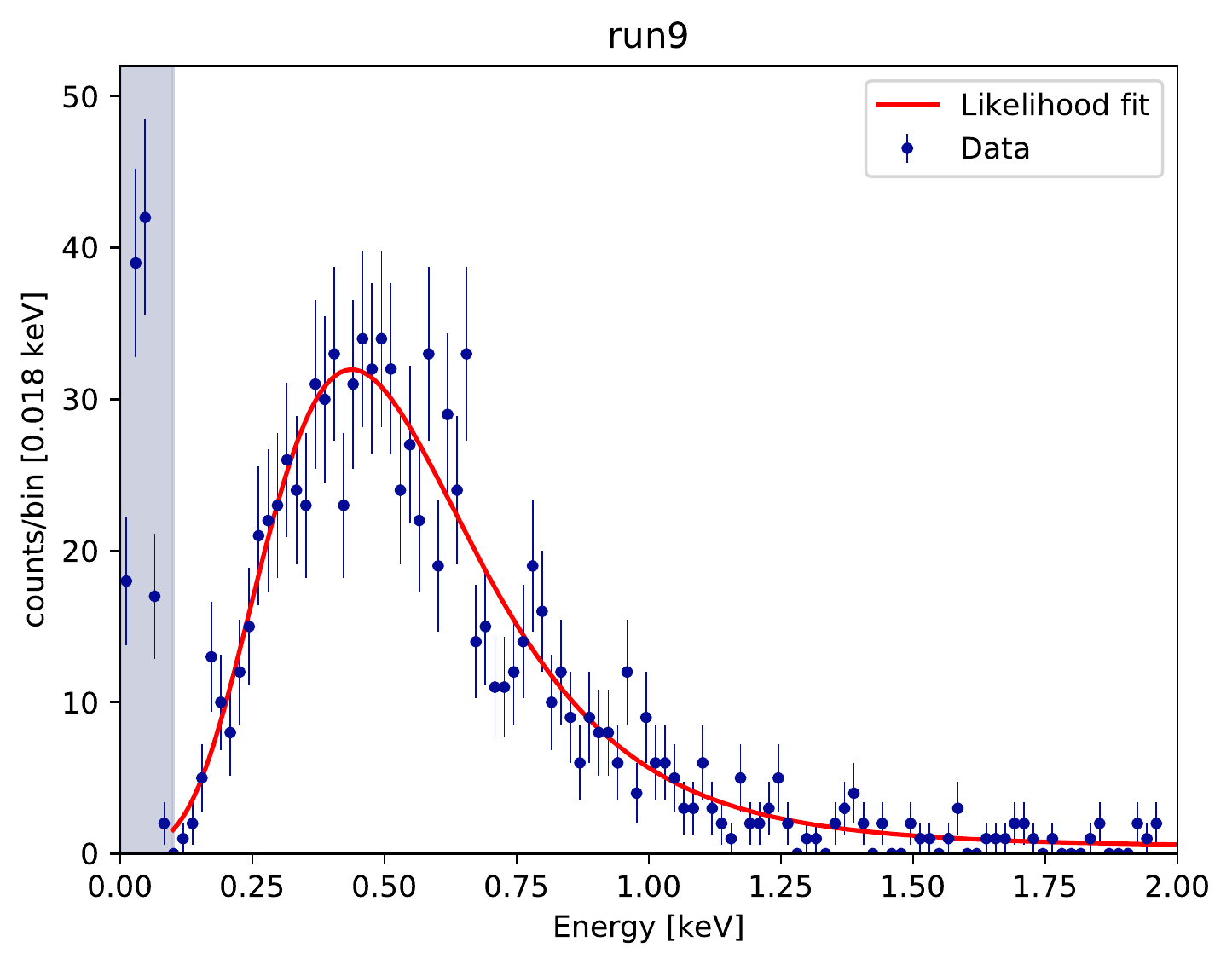}\quad
\includegraphics[width=.3\textwidth]{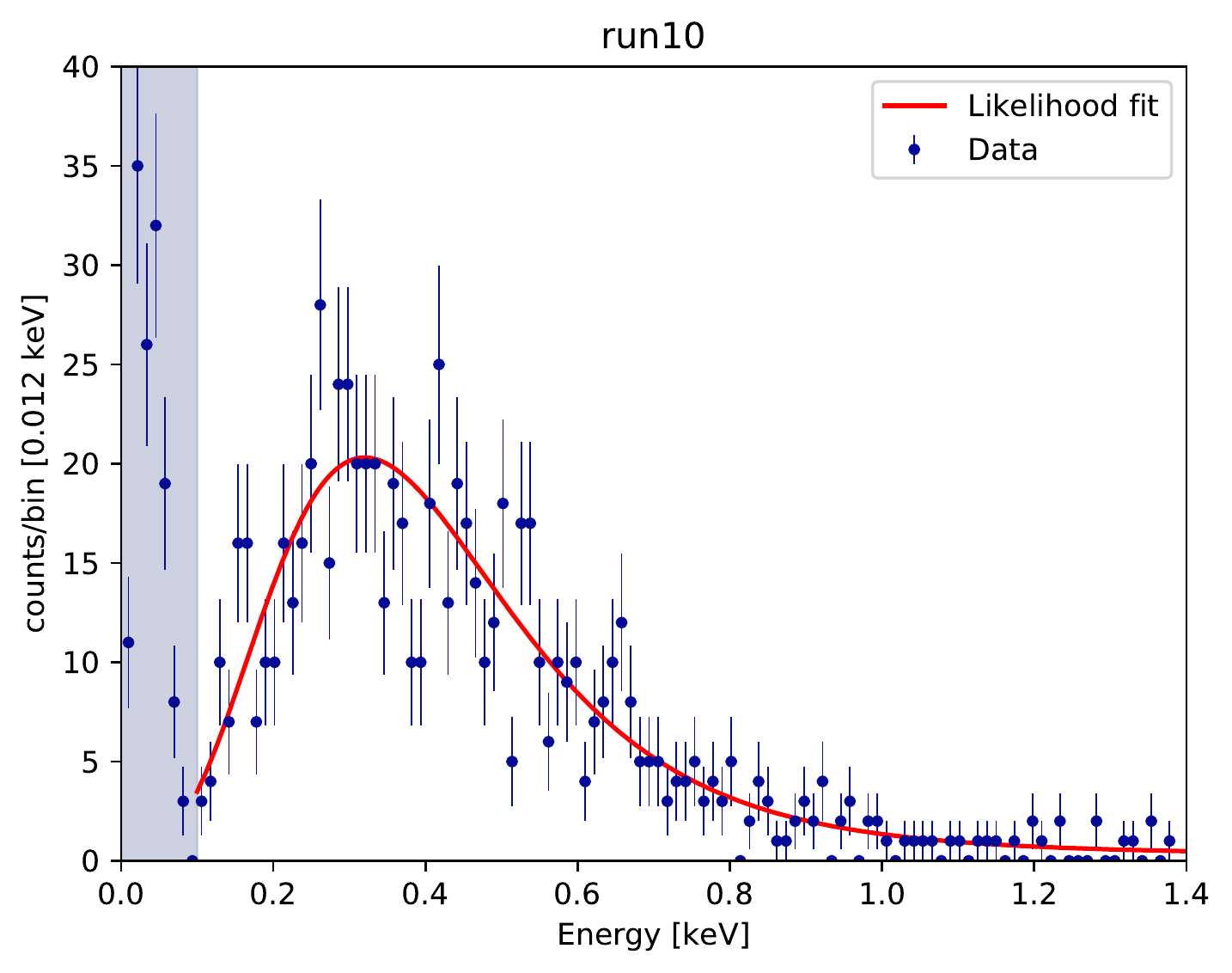}\quad
\includegraphics[width=.3\textwidth]{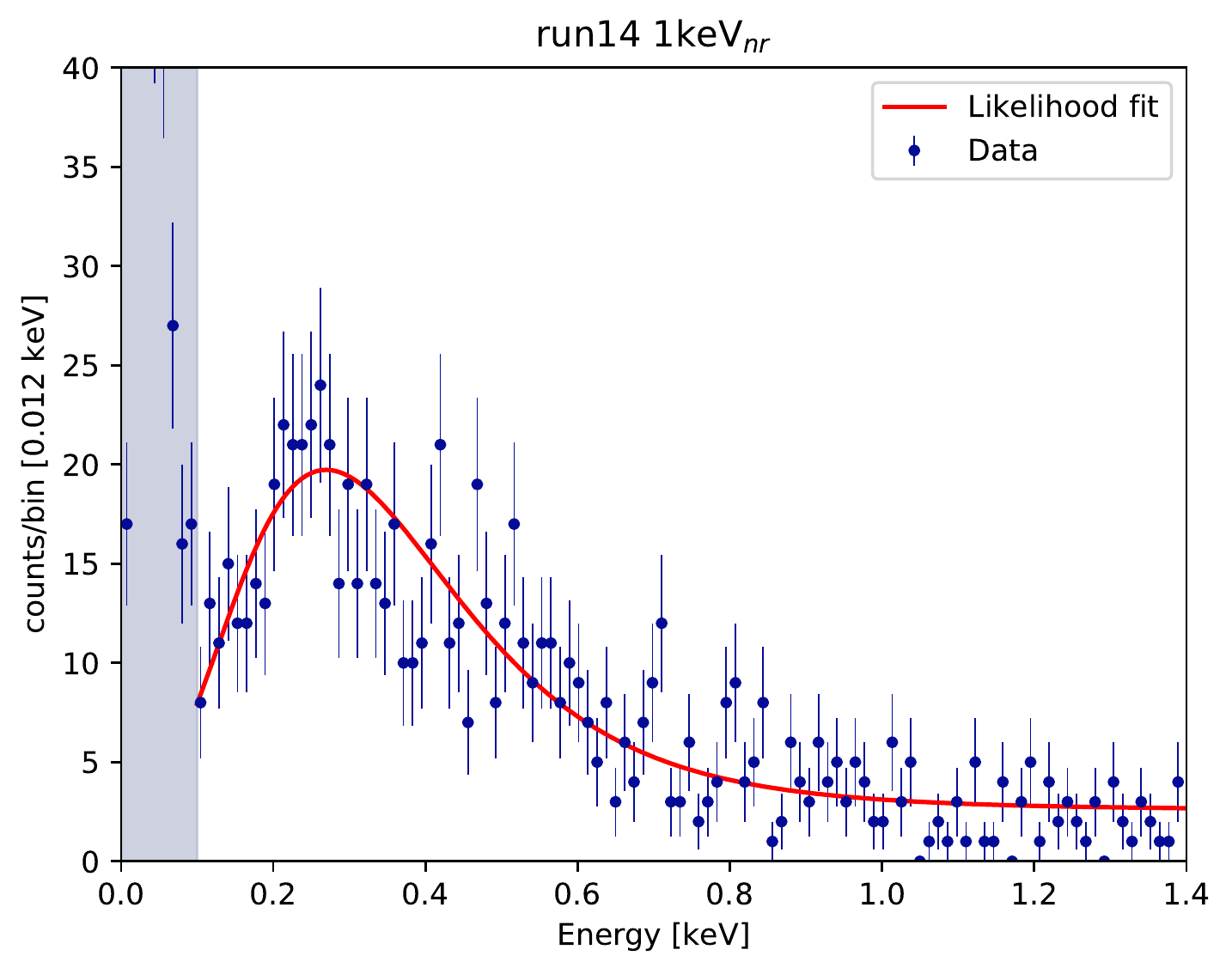}
\medskip
\includegraphics[width=.3\textwidth]{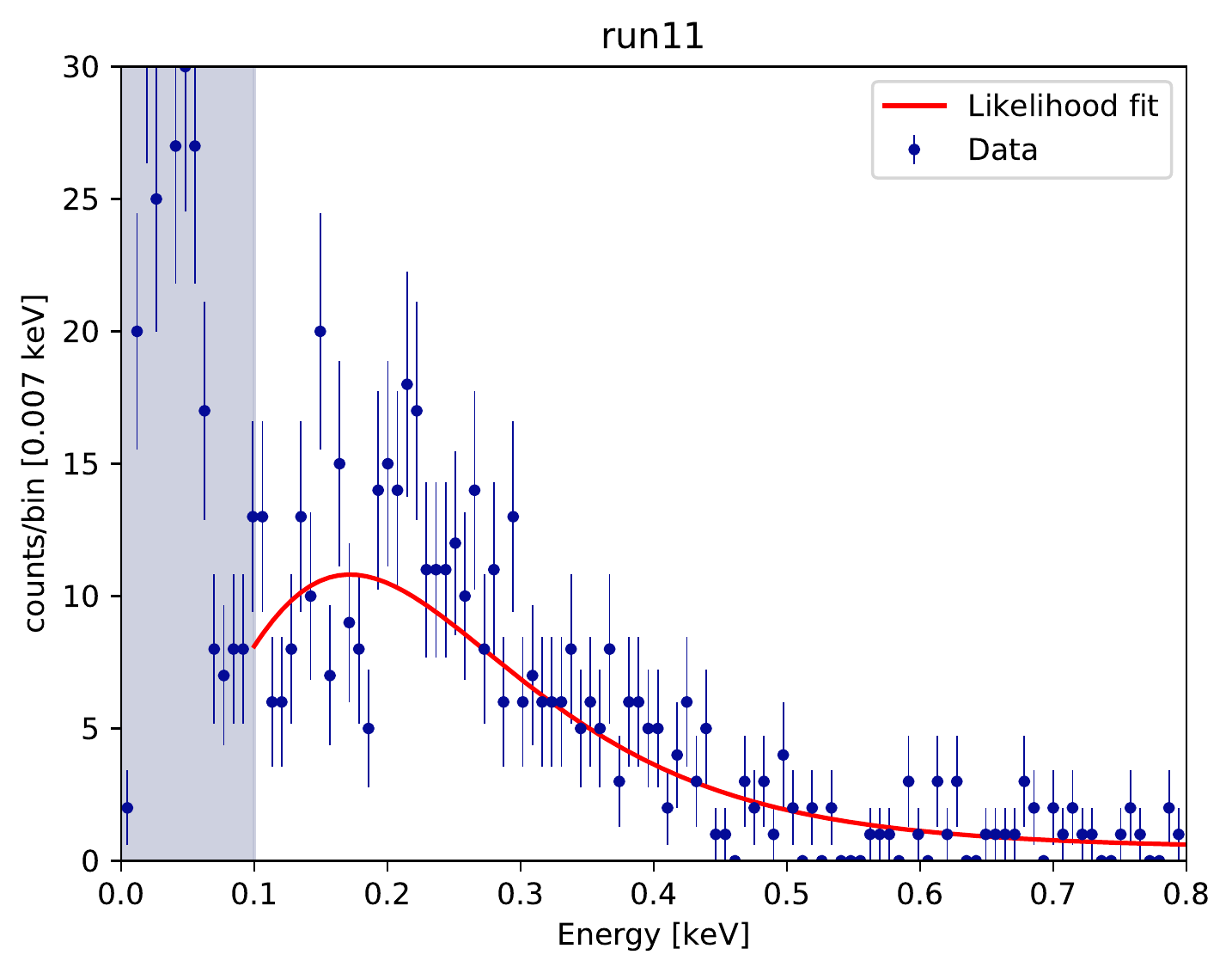}\quad
\includegraphics[width=.3\textwidth]{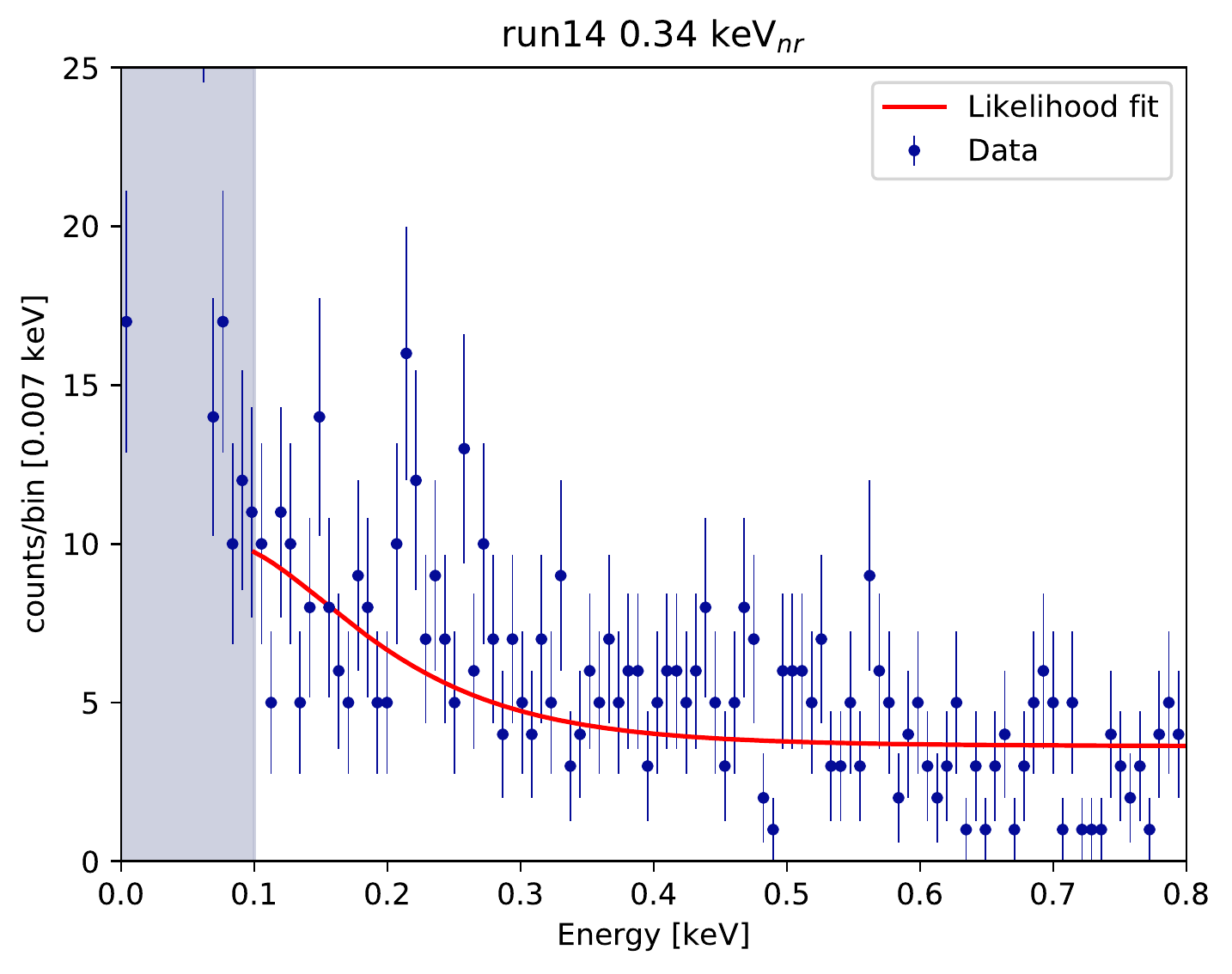}
\caption{\label{tab:results_fit} Multiple fit of all the data sets. In blue are the data arranged in a histogram and in red the fit (normalized to the data) with the most probable values for the free parameters returned by the minimizer. Recall that the mean gain, the W-value, $\theta_p$ and the neutron energy are fixed, and that the fraction of expected signal events for each run, as well as the scattering angles, the standard deviation of the energy scale, $\sigma_a$, $\alpha$ and $\beta$ are free parameters of the joint fit. The shaded area indicates the energy region below the analysis threshold.}
\end{figure*}

\section{Results and conclusion}\label{section4}

\subsection{Results}

In this work, we measured the nuclear recoil response of 2 bar of Ne + CH$_4$  (\SI{3}{\%}) in a SPC. The data were simultaneously fitted using the minimizer Minuit. The energy runs were fitted using different energy ranges. To model the recoil events the fit takes 19 free parameters, which include $\alpha$ and $\beta$, the parameters of the quenching factor function. The background is modelled by a uniform contribution in energy. 

Fig. \ref{tab:results_fit} shows the eight energy spectra, after all cuts were applied, with the results of the fit, which show good agreement for the most part of the energy runs except for run14 at \SI{2}{keV}. The Pearson's $\chi^2$/ndf and the P-value were calculated for each run, and are summarized in Table \ref{tab:table_chi}. A possible explanation for the poor fit in run14 at \SI{2}{keV_{nr}} could be the lack of precision which the scattering angles were determined for run14, as compared to the runs recorded with the annulus structure. Thus, when giving the prior on the scattering angles for run14, the error on the angle might have been underestimated and thus over-constrained in the fit. \\

The values returned for each free parameter are given in Table \ref{tab:final_res}. The expected fractions of signal events are consistent with each other for the runs taken with the annulus structure. The exposure time was adjusted so that we would record similar statistics in each run. For run14, the fraction of expected signal events decreases with the energy, due to the angular distribution of the scattered neutrons which favors \SI{90}{^o} angles, which is observed from previous experiments \cite{grayson}. \\

\begin{table}
\centering
	\begin{tabular}{ c c c c c c }
			\hline
			Run number & $\chi^2$ & ndf & $\chi^2$/ndf  & P-values & E$_{max}$ [keV]\\[0.5ex]
			\hline
			run 8  & 89.55 & 82 & 1.09 & 0.26 & 6\\
			run 7 & 67.46 & 80 & 0.84 & 0.84 & 3\\
			run 9 & 59.75 & 78 & 0.77 & 0.93 & 2 \\
			run 10 & 74.74 & 77 & 0.97 & 0.55 & 1.4\\
			run 11 & 65.79 & 81 & 0.81 & 0.89 & 0.8\\
			run 14:  \SI{2}{keV} & 140.4 & 87 & 1.69 & 2.5e-4 & 2.\\
			run 14:  \SI{1}{keV} & 92.75 & 89 & 1.04 & 0.37 & 1.4 \\
			run 14: \SI{0.34}{keV} & 88.75 & 87 & 1.02 & 0.43 & 0.8\\[1ex]
			\hline
		\end{tabular}
		\caption{Summary table of the $\chi^2$, number of degrees of freedom (ndf), $\chi^2$/ndf and p-values for the joint fit to all energy spectra. The last column gives the maximum energies of the energy ranges covered by the fit. \label{tab:table_chi}}
\end{table}

The scattering angle means, $\theta_s$, returned by Minuit are consistent within 1$\sigma$ error with the measurements, except for run14 at \SI{2}{keV} and \SI{1}{keV} and run8, which are at \SI{3.7}{\sigma}, \SI{1.4}{\sigma} and \SI{1.17}{\sigma} from their measured scattering angle respectively. \\
The standard deviation of the energy scale distribution is found to be \SI{23.8}{\%} in our model. This indicates non negligible fluctuation of the energy scale, or gain, throughout the volume of the detector we used. This may also be attributed to larger than Poissonian fluctuations in the NR recoil energy deposition process.\\

\begin{table}
	\begin{center}
		\begin{tabular}{ c c c c }
			\hline
			Parameters & Values & Uncertainties & Expected ($\theta_s$)\\ [0.5ex]
			\hline
			f$_{s8}$ & 0.922 & 0.020 & -\\
			f$_{s7}$ & 0.947 & 0.022&-\\
			f$_{s9}$ & 0.954 & 0.018&-\\
			f$_{s10}$ & 0.958 & 0.023&-\\
			f$_{s11}$ & 0.898 & 0.035&-\\
			f$_{s14, \SI{0.34}{keV}}$ & 0.234 & 0.039&-\\
			f$_{s14, \SI{1}{keV}}$ & 0.668 & 0.033&- \\
			f$_{s14, \SI{2}{keV}}$ & 0.733 & 0.029&-\\
			$\alpha$ & 0.2801 & 0.0050&-\\
			$\beta$ & 0.0867 & 0.020&-\\
			$\theta_8$ & 28.55 & 0.34& 29.02 $\pm$ 0.4\\
			$\theta_7$ & 18.89 & 0.09& 18.84 $\pm$ 0.1\\
			$\theta_9$ & 14.33 & 0.06& 14.33 $\pm$ 0.06\\
			$\theta_{10}$ & 12.47 & 0.048& 12.48 $\pm$ 0.05\\
			$\theta_{11}$ & 9.41 & 0.033 & 9.4 $\pm$ 0.03\\
			$\theta_{14, 0.34keV}$ & 6.57 & 0.26 & 6.33 $\pm$ 0.26\\
			$\theta_{14, 1keV}$ & 11.55 & 0.18 & 11.13 $\pm$ 0.3\\
			$\theta_{14, 2keV}$ & 14.51 & 0.15 & 15.63 $\pm$ 0.3\\
			$\sigma_a$ & 0.238 & 0.01&- \\[1ex]
			\hline
		\end{tabular}
	\caption{\label{tab:final_res} Table summarizing the most probable values of the free parameters returned by the fit and the expected values for the scattering angles, $\theta_s$ and their uncertainties. The eight $f_s$ are the expected fraction of signal events for each energy run. $\alpha$ and $\beta$ are the parameters of the quenching factor function. Finally, $\sigma_a$ is the standard deviation of the energy scale (gain) distribution.}
	\end{center}
\end{table}

For comparison, we also fit each run individually, we reported the modes of the energy runs from the annulus structure, which provided the best accuracy for the scattering angles, into the quenching factor as a function of $E_{nr}$ plot:
\begin{itemize}
    \item[--]run8: \SI{6.5}{keV_{nr}} 
    \item[--]run7: \SI{2.9}{keV_{nr}} 
    \item[--]run9: \SI{1.7}{keV_{nr}} 
    \item[--]run10: \SI{1.3}{keV_{nr}} 
    \item[--]run11: \SI{0.74}{keV_{nr}} 
\end{itemize}

\begin{figure*}
\centering
\includegraphics[width=.6\textwidth]{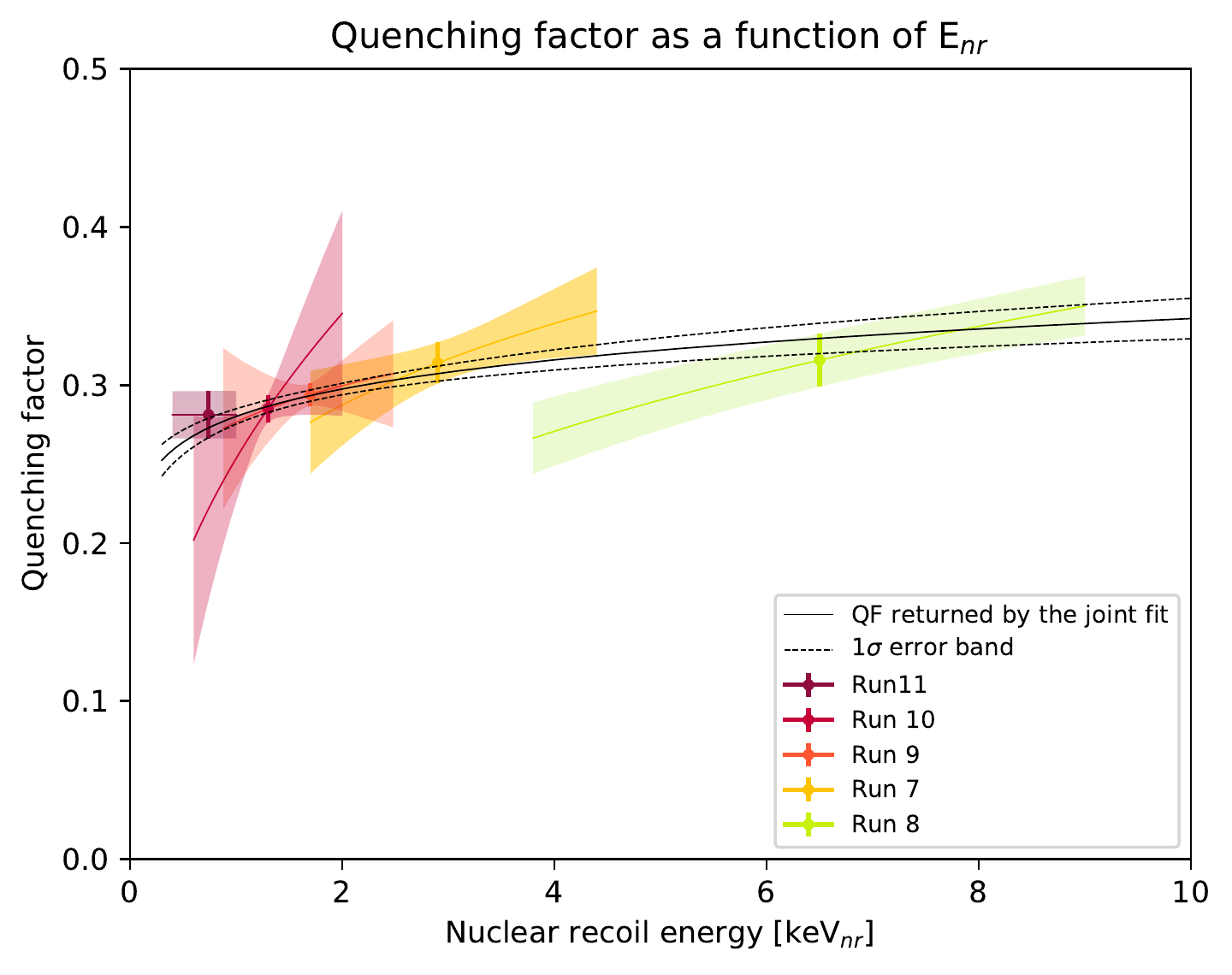}
\caption{Shows the individual QF, color curves, for the modes $\pm$\SI{1}{\sigma} of five of the energy runs (annulus structure), along with the joint fit result, solid black line, and \SI{1}{\sigma} error band, black dashed line.\label{tab:QF_ind_multi}}
\end{figure*}

Fig. \ref{tab:QF_ind_multi} shows the independent quenching factors for runs 7, 8, 9, 10 and 11, in color, as well as the quenching factor from the joint fit, in black. All error bands correspond to \SI{1}{\sigma} error including statistical errors and systematic errors from the scattering angles returned by the fit. All of the errors reported were calculated using the covariance matrix returned by Minuit. The results from the individual fits are consistent with the result from the joint fit, indicating a suitable parametrization to model the quenching factor. Fig. \ref{tab:QF_ind_multi} also highlights how the model that we used allows one to leverage multiple and overlapping energy ranges to provide a stronger constraint on the QF. The simultaneous fit over all energies also reduces the effect of the correlation between $\alpha$ and $\beta$ that would otherwise lead to large uncertainties (see for example error bands on runs 7, 9, and 10). The rather small uncertainties at the mode of the recoil energy distributions show that the size of the uncertainties is driven by the statistics available at a given energy. When fitting all the energy runs simultaneously, the region around \SI{2}{keV} has the most statistics, and is thus where we have the strongest constraints on the QF.\\

\indent A systematic uncertainty due to possible non-linearities in the energy scale was investigated by implementing a quadratic energy response. The quadratic term was fixed in such a way as to produce a maximum non-linearity of \SI{0.7}{\%} in the position of the \SI{2.82}{keV} peak from $^{37}$Ar when the $^{55}$Fe peak is held fixed. The \SI{0.7}{\%} maximal non-linearity would be consistent with data collected with a similar SPC filled with \SI{1.5}{bar} of Ne+CH$_4$ (\SI{2}{\%}) and evaluated with the \SI{270}{eV} and \SI{2.82}{keV} X-ray lines, and described in \cite{laser}. The impact of the non-linearity by $\pm$\SI{0.7}{\%} is at most of \SI{0.7}{\%} and \SI{0.6}{\%} on the quenching factor, at high (\SI{10}{keV_{nr}}) and at low energy (\SI{0.43}{keV_{nr}}), respectively. \\
\indent A systematic uncertainty due to the efficiency curve was implemented by changing the rise time cut value used to build such a curve. The value of the shift in the rise time cut was $\pm$ \SI{6}{\%}. The impact of such source of uncertainty is at most of \SI{0.29}{\%} and \SI{0.8}{\%} on the quenching factor, at high (\SI{10}{keV_{nr}}) and at low energy (\SI{0.43}{keV_{nr}}), respectively.\\  
\indent A systematic uncertainty due to possible electronic offset, resulting in an energy response not going through zero, was investigated. For this, we looked at the noise peak energy spectrum, without cuts, and its location relative to zero. We found that the mean of the noise peak was at \SI{-13.7}{ADU}. The impact of such offset is at most of \SI{0.06}{\%} and \SI{1}{\%} on the quenching factor, at high (\SI{10}{keV_{nr}}) and at low energy (\SI{0.43}{keV_{nr}}), respectively.\\
\indent A systematic uncertainty due to the energy scale uncertainty was investigated. The error on the mean of the $^{55}$Fe calibration was on average is \SI{1.35}{\%}. The impact of such uncertainty is at most \SI{1.24}{\%} and \SI{1.3}{\%} on the quenching factor, at high (\SI{10}{keV_{nr}}) and at low energy (\SI{0.43}{keV_{nr}}), respectively.\\
\indent Bias and pull tests of the analysis framework were performed on simulated data. These showed a small bias in the fitted values of +\SI{0.34}{\%}  for $\alpha$ and -\SI{1.6}{\%} for $\beta$. Since these biases are small, we report them as a systematic uncertainties. Their impact on the quenching factor is at most of \SI{0.07}{\%} and \SI{0.47}{\%}, at high (\SI{10}{keV_{nr}}) and at low energy (\SI{0.43}{keV_{nr}}), respectively.\\
\indent The different sources of uncertainties are listed in Table \ref{tab:unc_summary}, with their impact on the parameters $\alpha$ and $\beta$, as well as impact on the quenching factor. The different impacts on the QF were estimated by looking at the maximum differences between the QF with no systematic uncertainty and the QFs obtained from the different systematic uncertainties studies. \\

The quenching factor model tested a parametrization of the form: $QF(E_{nr})$ = $\alpha E_{nr}^{\beta}$. The values of $\alpha$ and $\beta$ with the errors reported by Minuit (1$\textsuperscript{st}$ error: statistical and systematic from the scattering angle: fit) and the systematic error from potential non-linearities of the detector response (2$\textsuperscript{nd}$ error: sys), from the efficiency curve, from the electronic offset, from the energy scale and potential biases in the analysis are:
\begin{equation*}
\begin{split}
&\alpha = 0.2801 \pm 0.0050 \: \text{(fit)} \pm 0.0045 \: \text{(sys)}\quad \textrm{and} \\
&\beta = 0.0867 \pm 0.020 \: \text{(fit)}\pm 0.0069 \:\text{(sys)}
\end{split}
\end{equation*}
They apply to an energy range between 0.43 and \SI{11}{keV_{nr}}.\\

\begin{table}[ht]
	\begin{center}
		\begin{tabular}{ c c c c c }
			\hline
			 &  $\alpha$ & $\beta$ & QF(HE) & QF(LE) \\ [0.5ex]
			\hline
			Noise &  \SI{0.04}{\%} & \SI{0.4}{\%} & Negligible & Negligible\\
			Efficiency curve &\SI{0.4}{\%} & \SI{3.7}{\%} & $\leqslant$\SI{0.80}{\%}& $\leqslant$\SI{0.29}{\%}\\
			Non-linearity & \SI{0.21}{\%} & \SI{4.6}{\%} & $\leqslant$\SI{0.60}{\%}& $\leqslant$\SI{0.70}{\%}\\
			Offset & \SI{0.21}{\%} & \SI{4.6}{\%} & $\leqslant$\SI{0.06}{\%}& $\leqslant$\SI{1}{\%}\\
			Energy scale & \SI{0.21}{\%} & \SI{4.6}{\%} & $\leqslant$\SI{1.24}{\%}& $\leqslant$\SI{1.3}{\%}\\
			Analysis bias & -\SI{0.34}{\%} & +\SI{1.6}{\%} & $\leqslant$\SI{0.07}{\%} & $\leqslant$\SI{0.47}{\%}\\
			Total & \SI{1.6}{\%} & \SI{6.9}{\%} & $\leqslant$\SI{2.9}{\%} & $\leqslant$\SI{1.1}{\%}\\[1ex]
			\hline
		\end{tabular}
	\caption{\label{tab:unc_summary} Table summarizing the uncertainties investigated for the analysis and their impact on $\alpha$, $\beta$, and impact on the quenching factor. The fluctuation of the baseline noise has a negligible impact on the QF. The total systematic uncertainty counts the uncertainties from the efficiency curve, the possible non-linearity, the electronic offset, the energy scale and the analysis bias. The last two columns evaluate the maximum impact on the quenching factor at high and low energies (HE: \SI{10}{keV_{nr}} and LE: \SI{0.43}{keV_{nr}}) from these sources of uncertainties considering the $\alpha$s and $\beta$s returned from the different studies. The last raw corresponds to the total uncertainties on $\alpha$, $\beta$ and QF by adding in quadrature the different systematic contributions from the efficiency curve, the possible non-linearity, the electronic offset, the energy scale and the analysis bias.}
	\end{center}
\end{table}

\subsection{Conclusion}

\begin{figure*}
\centering
\includegraphics[width=.6\textwidth]{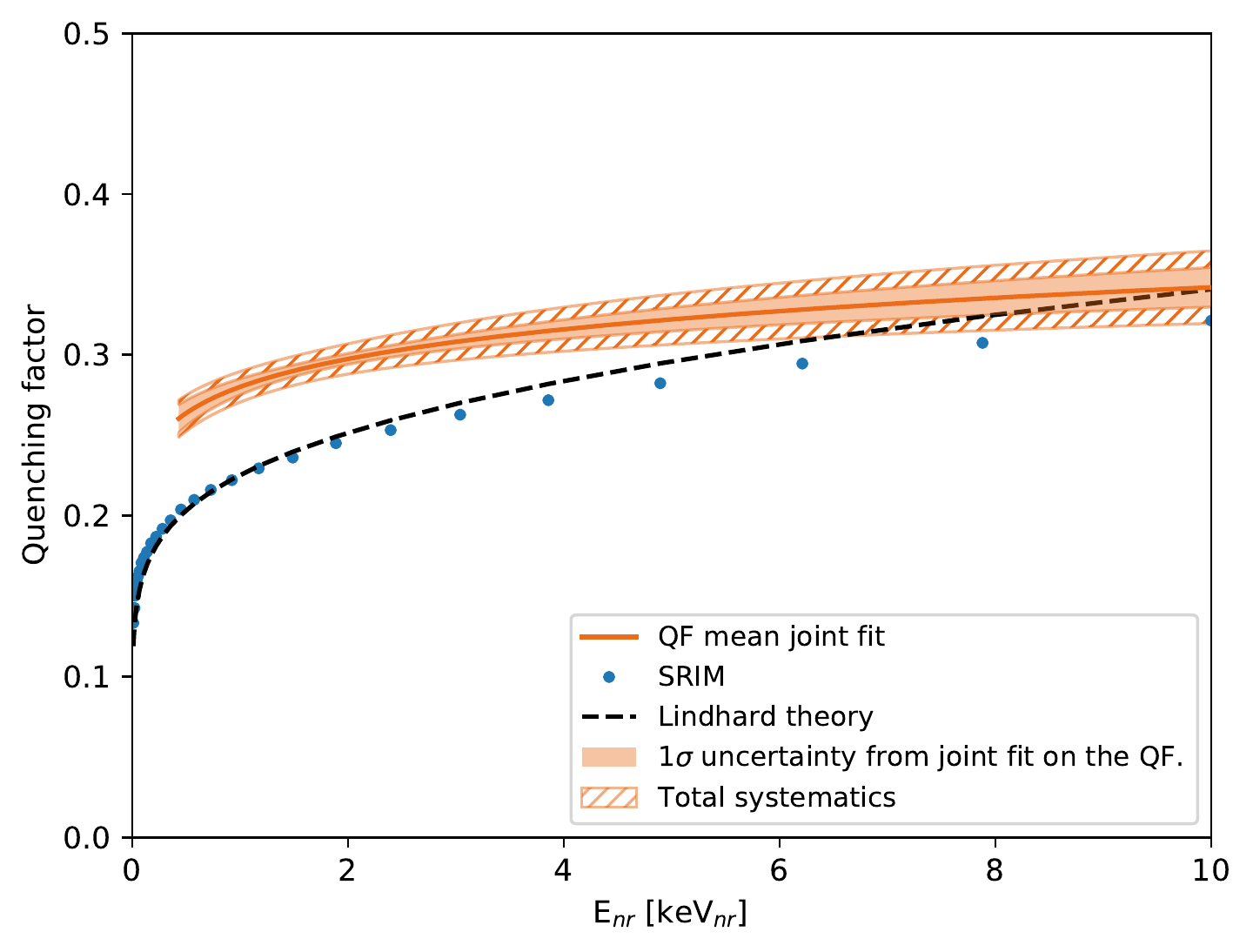}
\caption{Quenching factor as a function of the nuclear recoil energy, using the values of $\alpha$ and $\beta$ returned by the fit. The orange curve corresponds to the quenching factor mean for a given nuclear recoil energy. The orange error band corresponds to 1$\sigma$ error on the quenching factor for a given nuclear recoil energy. The dashed orange band corresponds to 1$\sigma$ systematic uncertainty on the quenching factor. The dashed black curve is the Lindhard theory and the blue dots are the SRIM QF. \label{tab:QF_res}}
\end{figure*}

Lindhard theory is often used to compare with experimental results. It shows reasonable agreement in silicon and germanium but also in LXe and LAr \cite{LUX}, \cite{NEST}. Two measurements in gases were performed in $^4$He and isobutane, \cite{santos} \cite{isobutane}. They showed some discrepancy between Lindhard and their experimental results as well as with the SRIM simulation. Fig. \ref{tab:QF_res} shows the resulting QF as a function of the nuclear recoil energy between 0.43 and \SI{10}{keV_{nr}} along with 1$\sigma$ uncertainty bands corresponding to errors from the fit and systematic errors, compared to the Lindhard theory and the SRIM simulation. The Lindhard theory and the SRIM simulation are consistent with each other, but not with our measurement. The quenching factor extracted from the experiment and analysis is larger than the theory below \SI{9}{keV_{nr}}. The maximum discrepancy between our quenching factor and Lindhard/SRIM is \SI{24}{\%} at low energy ($<$\SI{1}{keV_{nr}}). Our results show that the quenching factor in neon is more optimistic than expected, which allows us to to have increased sensitivity compared to what is expected from these models.\\

With this experiment we demonstrated the feasibility of measuring the quenching factor of gas mixtures using a Spherical Proportional Counter in a neutron beam below \SI{1}{keV}. To the best of our knowledge,  it is the first time that a quenching factor is extracted using a joint fit using the $\alpha\beta$ quenching factor parametrization, so that the quenching factor is known for each energy across the range covered: 0.43 and \SI{11}{keV_{nr}}. This is also the first time such measurements were performed in neon gas.\\

This work demonstrates the feasibility of performing measurements of nuclear ionization quenching factors using a spherical proportional counter. In the future, dedicated measurement campaigns can derive precise quenching factor measurements in neon and other gases. These future measurements will aim to reduce uncertainties related to the energy response of the SPC. This will primarily be achieved by including an electric field corrector to make the electric field, and hence the response, uniform in the detector volume. The uniform response of the detector will allow \textit{in situ} energy calibrations with gaseous $^{37}$Ar source, providing X-rays at \SI{270}{eV} and \SI{2.82}{keV} \cite{laser}, in addition to the \SI{5.9}{keV} calibration point that can be obtained from $^{55}$Fe as was used in this work. These additional 
energy calibration points will enable stricter control of a number of effects, for example any residual non-linearities in the energy response. Furthermore, future measurements will explore various gas mixtures and pressures in order to extend the reach of the NEWS-G physics programs.

\section{Acknowledgments}
\vspace{4.mm}
This research was undertaken, in part, thanks to funding from the Canada Excellence Research Chairs Program. We acknowledge the support of the Natural Sciences and Engineering Research Council of Canada, funding reference number SAPIN-2017-00023. We would like to thank the US Department of Energy (DE-FG02-97ER41033) for the use of the TUNL facility.

\section{Appendix}
We provide the covariance matrix to the reader, as it was used to calculate the errors and take into account any correlation between the different parameters, see Table \ref{tab:cov}. The correlation matrix is also provided, see Table \ref{tab:corr}.
\begin{table*}[!htbp]
\centering
\resizebox{\textwidth}{!}{\begin{tabular}{ c c c c c c c c c c c c c c c c c c c c}
		\hline
			Parameters & f$_{s8}$ & f$_{s7}$ &  f$_{s9}$ & f$_{s10}$ & f$_{s11}$ & f$_{s14, \SI{0.34}{keV}}$ & f$_{s14, \SI{1}{keV}}$ & f$_{s14, \SI{2}{keV}}$ & $\alpha$ & $\beta$ & $\theta_8$ & $\theta_7$ & $\theta_9$ & $\theta_{10}$ & $\theta_{11}$ & $\theta_{14, 0.34keV}$ & $\theta_{14, 1keV}$ & $\theta_{14, 2keV}$ & $\sigma_a$\\ [0.5ex]
			\hline
			f$_{s8}$ & 4.12e-4 & 7.65e-5 & 6.88e-5 & 6.98e-5 & 5.65e-5 & 1.99e-5 & 6.37e-5 & 8.07e-5 & -1.27e-5 & 3.38e-5 &  -4.67e-4 & -4.60e-5 & 8.60e-6 & 1.11e-5 & 2.86e-6&  -2.94e-6 &  8.68e-5 & -4.86e-5 & 7.03e-5\\
			f$_{s7}$ & 7.65e-5 & 4.91e-4 & 9.69e-5 & 1.02e-4 & 9.32e-5 & 3.68e-5 & 8.99e-5 & 1.09e-4 & -7.63e-6 & 1.01e-5 & -2.63e-4 & 8.20e-5  & -1.33e-5 & -4.04e-6 & -5.97e-7 & -3.79e-5 & -1.60e-6 & -1.49e-4 & 9.68e-5\\
			f$_{s9}$ & 6.87e-5 & 9.69e-5 &  3.42e-4 & 1.08e-4 & 1.11e-4 & 4.83e-5 & 9.15e-5 & 1.05e-4 & 4.03e-6 & -4.04e-5 & 3.23e-4 & -2.42e-5 & 6.64e-5 & -2.55e-5 & -6.18e-6 & -3.50e-5 & -1.43e-4 & -2.23e-4 & 9.61e-5\\
			f$_{s10}$ & 6.98e-5 & 1.025e-4 & 1.08e-4 & 5.40e-4 & 1.37e-4 & 6.21e-5 & 1.02e-4 & 1.13e-4 & 1.38e-5 & -8.25e-5&  7.73e-4 & -8.95e-6 & -6.24e-5 & 7.16e-5 & -1.12e-5 & -3.60e-5 & -2.71e-4 & -3.12e-4 & 1.05e-4\\
			f$_{s11}$ & 5.65e-5 & 9.32e-5 & 1.11e-4 & 1.36e-4 & 1.23e-3 & 8.41e-5 & 1.06e-4 & 1.08e-4 & 3.45e-5  & -1.78e-4 & 1.92e-3 & 6.20e-5 & -1.01e-4 & -8.32e-5 & 6.88e-5 & -3.18e-5 & -5.23e-4  &-4.29e-4 & 1.05e-4\\
			f$_{s14, \SI{0.34}{keV}}$ & 1.99e-5 & 3.67e-5 & 4.83e-5 & 6.21e-5 & 8.41e-5 & 1.53e-3 & 4.66e-5 & 4.45e-5 & 2.11e-5 & -1.07e-4 & 1.21e-3 & 5.29e-5 & -5.44e-5 & -4.68e-5 & -1.23e-5 & 2.94e-3 & -3.01e-4 & -2.15e-4 & 4.55e-5\\
			f$_{s14, \SI{1}{keV}}$ & 6.36e-5 & 8.99e-5 & 9.15e-5 & 1.02e-4 & 1.06e-4 & 4.66e-5 & 1.13e-3 & 9.91e-5 & 3.70e-6 & -4.68e-5 & 4.88e-4 & 3.80e-6 & -2.84e-5 & -2.23e-5 & -6.07e-6 & -3.38e-5 & 1.88e-3 & -1.82e-4 & 9.12e-5\\
			f$_{s14, \SI{2}{keV}}$ & 8.07e-5 & 1.09e-4 & 1.05e-4 & 1.13e-4 & 1.08e-4 & 4.45e-5 & 9.91e-5 & 8.67e-4 & -6.82e-6 & -1.48e-5 & 1.81e-4 & -1.13e-6 & -4.00e-6 & -4.10e-6 & -2.02e-6 & -4.29e-5 & -1.75e-5 & -3.00e-4 & 1.07e-4\\
			$\alpha$ & -1.26e-5 & -7.63e-6 & 4.04e-6 & 1.38e-5 & 3.45e-5 & 2.11e-5 & 3.70e-6 & -6.82e-6 & 2.48e-5 & -7.96e-5 & 6.71e-4 & -1.54e-5 & -7.20e-5 & -5.06e-5 & -1.10e-5 & -8.51e-6 & -3.07e-4 & -2.41e-4 & -3.13e-6\\
			$\beta$ & 3.38e-5 & 1.01e-5 & -4.04e-5 & -8.25e-5 & -1.78e-4 & -1.07e-4 & -4.68e-5 & -1.48e-5 & -7.96e-5 & 4.30e-4 & -5.61e-3 & -4.23e-4 & 1.21e-4 & 1.41e-4 & 4.22e-5 & 1.07e-6 & 9.91e-4 & 3.55e-4 & -3.95e-5\\
			$\theta_8$ & -4.67e-4 & -2.63e-4 & 3.23e-4 & 7.73e-4 & 1.92e-3 & 1.21e-3 & 4.88e-4 & 1.81e-4 & 6.72e-4 & -5.61e-3 & 1.15e-1 & 9.25e-3 & 3.72e-4 & -8.76e-4 & -4.23e-4 & -2.16e-4 & -8.12e-3 & 2.58e-3 & 5.78e-4\\
			$\theta_7$ & -4.60e-5 & 8.20e-5 & -2.42e-5 & -8.95e-6 & 6.20e-5 & 5.29e-5 & 3.80e-6 & -1.12e-6 & -1.54e-5 & -4.23e-4 & 9.25e-3 & 8.01e-3 & 3.85e-4 & 1.10e-4 & -8.49e-6 & -6.87e-5 & 2.65e-4 & 1.53e-3 & 4.37e-5\\
			$\theta_9$ & 8.60e-6 & -1.33e-5 & 6.64e-5 & -6.24e-5 & -1.02e-4 & -5.44e-5 & -2.84e-5&  -4.00e-6 & -7.20e-5 & 1.21e-4 & 3.73e-4 & 3.85e-4 & 3.06e-3 & 1.69e-4 & 2.91e-5 & -2.76e-6 & 9.26e-4 & 1.07e-3 & -1.09e-6\\
			$\theta_{10}$ & 1.10e-5 & -4.04e-6 & -2.55e-5 & 7.16e-5 & -8.32e-5 & -4.68e-5 & -2.23e-5 & -4.10e-6 & -5.06e-5 & 1.41e-4 & -8.76e-4 & 1.10e-4 & 1.69e-4 & 2.29e-3 & 2.23e-5 & 1.36e-5 & 6.44e-4 & 5.96e-4 & -7.40e-6\\
			$\theta_{11}$ & 2.86e-6 & -5.97e-7 & -6.18e-6 & -1.12e-5 & 6.88e-5 & -1.23e-5 & -6.07e-6 & -2.02e-6 & -1.10e-5 & 4.22e-5 & -4.23e-4 & -8.50e-5 & 2.90e-5 & 2.23e-5 & 1.13e-3 & 1.51e-6 & 1.39e-4 & 9.90e-5 & -3.74e-6\\
			$\theta_{14, 0.34keV}$ & -2.94e-5 & -3.79e-5 & -3.50e-5 & -3.60e-5 & -3.18e-5 & 2.94e-3 & -3.38e-5 & -4.29e-5 & 8.51e-6 & 1.07e-6 & -2.16e-4 & -4.87e-5 & -2.76e-5 & 1.36e-5 & -1.51e-6 & 6.87e-2 & -7.21e-2 & -6.38e-5 & -3.96e-5\\
			$\theta_{14, 1keV}$ & 8.68e-5 & -1.60e-6 & -1.43e-4 & -2.71e-4 & -5.23e-4 & -3.01e-4 & 1.88e-3 & -1.75e-5 & -3.07e-4 & 9.91e-4 & -8.12e-3 & 2.65e-4 & 9.27e-4 & 6.44e-4 & 1.39e-4 & -7.21e-5 & 3.21e-2 & 3.20e-3 & -5.18e-5\\
			$\theta_{14, 2keV}$ & -4.86e-5 & -1.49e-4 & -2.23e-4 & -3.12e-4 & -4.29e-4 & -2.15e-4 & -1.82e-4 & 3.00e-4 & -2.41e-4 & 3.55e-4 & 2.58e-3 & 1.52e-3 & 1.07e-3 & 5.95e-4 & 9.90e-5 & -6.38e-5 & 3.20e-3 & 2.32e-2 & -9.08e-5\\
			$\sigma_a$ & 7.03e-5 & 9.68e-5 & 9.61e-5 & 1.05e-4 & 1.06e-4 & 4.55e-5 & 9.12e-5 & 1.07e-4 & -3.13e-6 & -3.94e-5 & 5.78e-4 & 4.37e-5 & -1.1e-6 & -7.40e-6 & -3.74e-6 & -3.96e-5 & -5.18e-5 & -9.08e-5 & 9.94e-5\\[1ex]			\hline
	\end{tabular}}
	\caption{\label{tab:cov}  Covariance matrix for the joint fit performed to the data, provided by Minuit.}
	\end{table*}

\begin{table*}
\centering
\resizebox{\textwidth}{!}{\begin{tabular}{ c c c c c c c c c c c c c c c c c c c c}
		\hline
		Parameters & f$_{s8}$ & f$_{s7}$ &  f$_{s9}$ & f$_{s10}$ & f$_{s11}$ & f$_{s14, \SI{0.34}{keV}}$ & f$_{s14, \SI{1}{keV}}$ & f$_{s14, \SI{2}{keV}}$ & $\alpha$ & $\beta$ & $\theta_8$ & $\theta_7$ & $\theta_9$ & $\theta_{10}$ & $\theta_{11}$ & $\theta_{14, 0.34keV}$ & $\theta_{14, 1keV}$ & $\theta_{14, 2keV}$ & $\sigma_a$\\ [0.5ex]
			\hline	
	f$_{s8}$ & 1.000     &   0.170   &     0.183   &   0.148   &   0.079   &   0.025  &    0.093    &      0.135    &     -0.125    &      0.080     &   -0.068     &    -0.025     &     0.008    &     0.011      &    0.004     &    -0.006     &     0.024    &     -0.016    &      0.347\\ 
        f$_{s7}$& 0.170  &  1.000  &    0.236   &     0.199     &     0.120     &     0.042     &     0.121     &     0.168   &      -0.069     &     0.022     &    -0.035    &      0.041    &     -0.011     &    -0.004    &     -0.001    &     -0.007     &    -0.000    &     -0.044      &    0.438 \\
         f$_{s9}$ &0.183   & 0.236   &   1.000 &    0.252    &      0.171     &     0.067    &      0.147     &     0.194    &      0.044   &      -0.105    &      0.051     &    -0.015     &     0.065   &  -0.029    &     -0.010    &     -0.007   &      -0.043   &      -0.079     &     0.521 \\
        f$_{s10}$& 0.148   &   0.199     &     0.252  &  1.000      &    0.168    &  0.068     &     0.131     &     0.165      &    0.119     &    -0.171     &     0.098   &      -0.004      &   -0.048   &  0.064    &     -0.014    &     -0.006     &    -0.065    &     -0.088     &     0.453 \\
        f$_{s11}$ & 0.079 &  0.120 &  0.171     &     0.168     &     1.000    &      0.061      &    0.090    &      0.105     &     0.197    &     -0.244    &      0.161    &      0.020     &    -0.052   &  -0.050   &     0.058     &    -0.003    &     -0.083    &     -0.080    &      0.302 \\
        f$_{s14, 0.34keV}$&   0.025 & 0.042  &  0.067  &  0.068&0.061     &     1.000     &     0.035      &    0.039     &     0.108    &     -0.132     &     0.091     &     0.015     &    -0.025    &     -0.025 &  -0.009  &     0.286    &     -0.043    &     -0.036    &      0.117 \\
          f$_{s14, 1keV}$ &0.093  &  0.121   &  0.147  &  0.131   & 0.090 & 0.035     &     1.000    &      0.100     &     0.022    &     -0.067      &    0.043    &      0.001     &    -0.015    &     -0.014     &    -0.005     &    -0.004      &    0.312    &     -0.036     &     0.272 \\
          f$_{s14, 2keV}$ & 0.135  & 0.168  &  0.194  &   0.165  &0.105  &  0.039     &     0.100    &      1.000    &     -0.046     &    -0.024      &    0.018    &     -0.000     &    -0.002    &     -0.003    &     -0.002     &    -0.006     &    -0.003    &      0.067    &      0.367 \\
        $\alpha$ &-0.125    &     -0.069     &     0.044     &     0.119      &    0.197      &    0.108      &    0.022    &     -0.046     &     1.000     &    -0.770    &      0.397     &    -0.035      &   -0.261      &   -0.212     &    -0.066      &    0.007     &    -0.343     &    -0.317     &    -0.063 \\
        $\beta$ & 0.080    &      0.022     &    -0.105     &    -0.171     &    -0.244    &     -0.132    &     -0.067    &     -0.024   &      -0.770      &    1.000     &    -0.797     &    -0.228     &     0.105    &      0.142    &      0.060    &      0.000    &      0.266     &     0.112    &     -0.191 \\
         $\theta_8$ &-0.068     &    -0.035    &      0.051    &      0.098     &     0.161     &     0.091     &     0.043    &      0.018    &      0.397   &      -0.797      &    1.000     &     0.305      &    0.020      &   -0.054    & -0.037    &     -0.002     &    -0.133     &     0.050     &     0.171\\ 
         $\theta_7$ &-0.025     &     0.041     &    -0.015    &     -0.004     &     0.020      &    0.015      &    0.001     &    -0.000     &    -0.035     &    -0.228     &     0.305    &      1.000     &     0.078    &      0.026    &  -0.003   &      -0.002     &     0.017     &     0.112    &      0.049 \\
          $\theta_9$ &0.008   &      -0.011    &      0.065     &    -0.048      &   -0.052    &     -0.025   &      -0.015     &    -0.002    &     -0.261     &     0.105      &    0.020     &     0.078     &     1.000     &     0.064    &   0.016     &    -0.002     &     0.093     &     0.127     &    -0.002 \\
          $\theta_{10}$ &0.011     &    -0.004    &     -0.029     &     0.064     &    -0.050     &    -0.025     &    -0.014     &    -0.003     &    -0.212    &     0.142     &    -0.054     &     0.026       &   0.064    &      1.000     &  0.014      &   -0.001     &     0.075      &    0.082     &    -0.016 \\
          $\theta_{11}$ &0.004     &    -0.001    &     -0.010    &     -0.014    &      0.058    &     -0.009     &    -0.005    &     -0.002     &    -0.066    &      0.060    &     -0.037   &      -0.003     &     0.016     &     0.014     &  1.000    &     -0.000    &      0.023     &     0.019    &     -0.011 \\
         $\theta_{14, 0.34keV}$ &-0.006    &     -0.007     &    -0.007    &     -0.006    &     -0.003     &     0.286     &    -0.004     &    -0.006     &     0.007     &     0.000    &     -0.002    &     -0.002    &     -0.002   &   -0.001  &    -0.000     &     1.000    &     -0.002    &  -0.002     &  -0.015 \\
         $\theta_{14, 1keV}$ & 0.024     &    -0.000     &    -0.043     &    -0.065    &     -0.083      &   -0.043     &     0.312     &    -0.003    &     -0.343     &     0.266    &     -0.133     &     0.017    &      0.093     &     0.075     &     0.023     &    -0.002    &      1.000    &  0.117      & -0.029 \\
         $\theta_{14, 2keV}$ &-0.016    &     -0.044     &    -0.079     &    -0.088     &    -0.080      &   -0.036    &     -0.036     &     0.067    &     -0.317     &     0.112    &      0.050     &     0.112      &    0.127      &    0.082   &       0.019     &    -0.002     &     0.117    &  1.000   &   -0.060 \\
          $\sigma_a$ &0.347     &     0.438      &    0.521     &     0.453      &    0.302     &     0.117      &    0.272       &   0.367      &   -0.063    &     -0.191      &    0.171     &     0.049      &   -0.002    &     -0.016      &   -0.011    &     -0.015     &    -0.029      &   -0.060  & 1.000 \\[1ex]			
          \hline
	\end{tabular}}
	\caption{\label{tab:corr} Correlation matrix for the joint fit performed to the data.}
	\end{table*}

\end{document}